\documentclass{article}

\usepackage[english]{babel}

\usepackage[a4paper,top=2cm,bottom=2cm,left=3cm,right=3cm,marginparwidth=1.75cm]{geometry}

\usepackage{amsmath, amssymb}
\usepackage{siunitx}
\PassOptionsToPackage{hyphens}{url}\usepackage{hyperref}
\usepackage{cleveref}
\usepackage[utf8]{inputenc}
\usepackage{csquotes}
\usepackage{booktabs}
\usepackage{longtable}
\usepackage{adjustbox}
\usepackage{array}
\usepackage{url}
\usepackage{titlesec}
\usepackage{authblk}
\usepackage{xcolor} 

\usepackage{subcaption}
\definecolor{myc1}{HTML}{003049}
\definecolor{myc2}{HTML}{d62828}
\definecolor{myc3}{HTML}{f77f00}
\definecolor{myc4}{HTML}{6ca13b}

\definecolor{Q2}{HTML}{1f77b4}
\definecolor{Q1}{HTML}{ff7f0e}
\definecolor{Q3}{HTML}{2ca02c}

\definecolor{Q2}{HTML}{6ca13b}
\definecolor{Q1}{HTML}{6ca13b}
\definecolor{Q3}{HTML}{6ca13b}

\usepackage{tikz}
\usepackage{tikz-3dplot}
\usepackage{pgfplots}
\usetikzlibrary{decorations.markings,arrows.meta,calc,shapes,intersections,backgrounds,patterns.meta,bending,angles,decorations.pathreplacing,decorations.pathmorphing,pgfplots.colorbrewer,math,positioning}
\usepgfplotslibrary{fillbetween}
\usepackage{xifthen}
\usepackage[normalem]{ulem}
\usepackage{xcolor}

\let\bs\boldsymbol
\newcommand{\bq}{\bs{q}}        
\newcommand{\bi}{{\bs{i}}}        
\newcommand{\bQ}{\bs{Q}}        
\newcommand{\elb}{\mathcal{E}_{\rm{LB}}}        
\newcommand{\xz}{\bs{x}_{\|}}        
\newcommand{\Is}{I}        
\newcommand{\GIs}{\tilde{\mathcal{T}}}  
\newcommand{\bj}{\bs{j}}        

\newcommand{\appropto}{\mathrel{\vcenter{
  \offinterlineskip\halign{\hfil$##$\cr
    \propto\cr\noalign{\kern2pt}\sim\cr\noalign{\kern-2pt}}}}}

\titleformat{\subsection}
  {\mdseries\itshape\large} 
  {\thesubsection}{1em}{} 

\usepackage[english]{babel}
\usepackage[style=authoryear,backend=biber,natbib=true,maxcitenames=2,uniquelist=false,dashed=false]{biblatex}
\DeclareNameAlias{sortname}{family-given}
\DeclareNameAlias{default}{family-given}

\renewbibmacro{in:}{}
\DeclareFieldFormat[article]{title}{\mkbibquote{#1}\addcomma}
\DeclareFieldFormat[book]{title}{\mkbibemph{#1}\addcomma}
\DeclareFieldFormat[bookinbook]{title}{\mkbibemph{#1}\addcomma}
\DeclareFieldFormat[inbook]{title}{\mkbibquote{#1}\addcomma}
\DeclareFieldFormat[incollection]{title}{\mkbibquote{#1}\addcomma}
\DeclareFieldFormat[inproceedings]{title}{\mkbibquote{#1}\addcomma}
\DeclareFieldFormat[manual]{title}{\mkbibemph{#1}\addcomma}
\DeclareFieldFormat[misc]{title}{\mkbibemph{#1}\addcomma}
\DeclareFieldFormat[thesis]{title}{\mkbibemph{#1}\addcomma}
\DeclareFieldFormat[unpublished]{title}{\mkbibquote{#1}\addcomma}
\DeclareFieldFormat[patent]{title}{\mkbibemph{#1}\addcomma}
\DeclareFieldFormat[report]{title}{\mkbibemph{#1}\addcomma}
\DeclareFieldFormat[online]{title}{\mkbibquote{#1}\addcomma}
\DeclareFieldFormat[software]{title}{\mkbibemph{#1}\addcomma}
\DeclareFieldFormat[booklet]{title}{\mkbibemph{#1}\addcomma}
\DeclareFieldFormat[periodical]{title}{\mkbibemph{#1}\addcomma}
\DeclareFieldFormat[standard]{title}{\mkbibemph{#1}\addcomma}

\DeclareFieldFormat[article]{journaltitle}{\iffieldundef{shortjournal}{\mkbibemph{#1}\addcomma}{\mkbibemph{\printfield{shortjournal}}\addcomma}}
\DeclareFieldFormat{volume}{\bibstring{volume}~#1}
\DeclareFieldFormat{number}{\bibstring{number}~#1}

\DefineBibliographyStrings{english}{
  volume = {Vol.},
  number = {No.}
}

\renewbibmacro*{volume+number+eid}{%
  \printfield{volume}%
  \setunit*{\addspace}%
  \printfield{number}%
  \setunit{\addcomma\space}%
  \printfield{eid}}

\renewbibmacro*{journal+issuetitle}{%
  \usebibmacro{journal}%
  \setunit*{\addcomma\space}%
  \usebibmacro{volume+number+eid}%
  \setunit{\addcomma\space}%
  \usebibmacro{issue+date}}

\renewbibmacro*{publisher+location+date}{%
  \printlist{publisher}%
  \iflistundef{location}
    {\setunit*{\addcomma\space}}
    {\setunit*{\addcolon\space}}%
  \printlist{location}%
  \setunit*{\addcomma\space}%
  \usebibmacro{date}}

\DeclareCiteCommand{\cite}[\mkbibparens]
  {\usebibmacro{prenote}}
  {\usebibmacro{citeindex}%
   \usebibmacro{cite}}
  {\multicitedelim}
  {\usebibmacro{postnote}}

\renewbibmacro*{cite:labelyear+extrayear}{%
  \iffieldundef{labelyear}
    {}
    {\printtext[bibhyperref]{%
       \printfield{labelyear}%
       \printfield{extrayear}}}}

\renewbibmacro*{cite:labeldate+extradate}{%
  \iffieldundef{labelyear}
    {}
    {\printtext[bibhyperref]{%
       \printfield{labelyear}%
       \printfield{extradate}}}}

\AtEveryBibitem{
  \clearfield{month}
  \clearfield{day}
  \ifentrytype{book}{
    \clearlist{location}
  }{}
}

\DefineBibliographyStrings{english}{
  andothers = {\textit{et al.},}
}

\DeclareFieldFormat[article]{volume}{\bibstring{jourvol}\addnbspace #1}
\DeclareFieldFormat[article]{number}{\bibstring{number}\addnbspace #1}
\DeclareFieldFormat[article]{volume}{Vol. #1}
\DeclareFieldFormat[article]{number}{No. #1}

\DeclareFieldFormat{url}{\bibstring{available at}\addcolon\space\url{#1}}
\DeclareFieldFormat{urldate}{\mkbibparens{accessed \addspace#1}}

\DeclareFieldFormat{urldate}{%
  \mkbibparens{accessed\space%
    \thefield{urlday}\addspace%
    \mkbibmonth{\thefield{urlmonth}}\addspace%
    \thefield{urlyear}}}

\crefformat{figure}{#2Figure~#1#3}
\Crefformat{figure}{#2Figure~#1#3}
\crefformat{table}{#2Table~#1#3}
\Crefformat{table}{#2Table~#1#3}
\crefformat{section}{#2Section~#1#3}
\Crefformat{section}{#2Section~#1#3}

\author[1]{Álvaro Martínez-Sánchez}
\author[1,2]{Adrián Lozano-Durán}

\affil[1]{Massachusetts Institute of Technology, Department of Aeronautics and Astronautics, Cambridge, USA
\href{mailto:alvaro@mit.edu}{alvaroms@mit.edu} (corr. author)}
\affil[2]{California Institute of Technology, Graduate Aerospace Laboratories, Pasadena, USA}

\title{Cause-and-effect approach to turbulence forecasting}

\begin{document}
\maketitle


\begin{abstract}
  \textbf{Purpose} – Traditional modeling techniques for forecasting
  turbulence often rely on correlation-based criteria, which may
  select variables that correlate with the target without truly
  driving its dynamics. This limits model interpretability,
  generalization, and efficiency. The purpose of this study is to
  overcome these limitations by introducing an observational
  causality-based approach to input selection that identifies the
  variables responsible for the future evolution of a target quantity
  while disregarding non-causal factors.

  \textbf{Design/Methodology/Approach} – Our approach is grounded in
  the Synergistic--Unique--Redundant Decomposition (SURD) of
  causality, which dissects the information that candidate inputs
  provide about a target variable into unique, redundant, and
  synergistic causal components. These components are directly linked
  to the theoretical limits of predictive performance, quantified
  through the information-theoretic notion of irreducible error. To
  estimate these causal contributions in practice, we leverage neural
  mutual information estimators. We demonstrate the methodology by
  forecasting wall-shear stress using direct numerical simulation
  (DNS) data of turbulent channel flow.

  \textbf{Findings} – The analysis reveals that variables with high
  unique or synergistic causal contributions enable compact
  forecasting models with strong predictive performance, whereas
  redundant variables can be excluded without compromising
  accuracy. Specifically, when predicting future wall-shear stress
  using two wall-parallel planes separated in the wall-normal
  direction, the streamwise velocity near the wall provides unique
  information about the target. In contrast, when both planes are
  located close to the wall, their information is largely redundant,
  and either can serve as input without degrading predictive
  accuracy. Finally, synergistic interactions emerge between different
  velocity components, which, when combined, enhance the prediction of
  future wall-shear stress beyond what each component achieves
  individually.

  \textbf{Originality/Value} – This work presents a causality-based
  approach for input selection in turbulence forecasting. The method
  quantifies the causal contributions of candidate variables to the
  prediction of a future quantity of interest and connects them to the
  fundamental limits of predictive accuracy achievable by any
  model. This enables more interpretable and compact models by
  reducing input dimensionality without sacrificing
  performance. Beyond turbulence, the approach provides a
  general-purpose tool for variable selection in scientific machine
  learning, flow control, and data-driven modeling of complex systems.
\end{abstract}


\section{Introduction}
\label{sec:introduction}

Among the physical sciences, fluid mechanics is distinguished by the
fact that its fundamental equations of motion---the Navier--Stokes
equations---are known and reproduce flow physics with remarkable
precision. Yet, despite this advantage, predicting turbulent flows
remains one of the most challenging problems in engineering and
scientific applications. The difficulty arises from the nonlinear and
multiscale nature of turbulence, which gives rise to a vast number of
interacting degrees of freedom. Capturing these dynamics directly from
the governing equations is computationally prohibitive for most
practical applications, motivating the development of reduced-order
models (ROMs) that retain the essential physics while reducing
dimensionality.

Over the past decades, many techniques have been developed to
construct such models. Classical approaches include Proper Orthogonal
Decomposition (POD) with Galerkin
projection~\cite{lumley1967,holmes2012}, balanced
truncation~\cite{moore1981}, and Dynamic Mode Decomposition
(DMD)~\cite{schmid2010}, as well as extensions based on Koopman
theory~\cite{edmd2015}. More recently, machine-learning methods have
entered the field, offering data-driven frameworks for model
construction~\cite{brunton2020}.  Applications of these techniques in
turbulence modeling are found in Reynolds-Averaged Navier--Stokes
(RANS) models (e.g., \citeauthor{ling2016}, \citeyear{ling2016}) and
Large-Eddy Simulation (LES) models (e.g., \citeauthor{arranz2024les},
\citeyear{arranz2024les}). These approaches reduce dimensionality by
not resolving all turbulent scales and introduce closure models to
represent the influence of unresolved motions on the resolved flow
variables. The development of such models is typically guided by
theoretical considerations, invariance principles, or empirical
fits~\cite{yuan2025}.  However, despite steady progress and the
promise of emerging data-driven techniques, the current generation of
models remains unable to meet the stringent accuracy and efficiency
demands of many scientific and industrial applications.

A fundamental challenge underlying these approaches is the selection
of input variables on which the models should be built. Effective
forecasting depends on identifying a minimal set of features that
offers a parsimonious yet sufficiently informative representation of
the system~\cite{guyon2003}. In practice, this is rarely
straightforward: turbulent flows involve many interacting features
across scales, blurring the distinction between variables that truly
drive the dynamics and those that merely correlate with
them~\cite{duraisamy2019, lozano2022, martinez2023, NatCom2024,
  arranz2024}.  For example, in aeronautics, one may wish to forecast
aerodynamic forces using limited measurements, such as velocity or
pressure at accessible locations. Using the entire flow field would
result in models of prohibitive complexity, while discarding too many
variables risks omitting the actual drivers. The central challenge,
therefore, is to identify the minimal and most informative set of
inputs that preserves the predictive content of the full system.

Traditionally, the selection of input variables has relied heavily on
heuristics and domain knowledge rather than rigorous principles, yet
it remains a critical step in building predictive models of
turbulence. Early efforts focused on filter
methods---see~\citet{biswas2016} for a review---which assess the
statistical dependence between individual features or groups of
features and the target variable~\cite{duch2003}. Examples include
correlation measures~\cite{mo2011}, fractal dimension~\cite{mo2010},
and distance measures~\cite{bins2001}.  Information theory has also
provided a rich foundation for these techniques, ranging from early
model selection criteria such as Akaike’s Information
Criterion~\cite{akaike1974} to modern information-theoretic methods
for coarse-graining and dynamical reduction~\cite{anderson2004,
  lozano2022, yuan2025}, as well as mutual information-based
approaches~\cite{meyer2008}, which aim to identify features that
maximize predictive association.

These methods are typically applied to individual variables and
therefore fail to capture the diverse types of interactions among
features. As a result, they cannot distinguish between variables that
are only informative when considered jointly (synergy) and those that
provide overlapping information about the target (redundancy). To
address this limitation, wrapper methods perform an iterative search
in which subsets of features are evaluated based on predictive
performance~\cite{kohavi1997, guyon2003}.  Approaches such as
Sequential Forward Selection (SFS)~\cite{whitney1971} and Sequential
Backward Elimination (SBE)~\cite{marill1963} progressively add or
remove variables to identify combinations that yield the most accurate
predictions. While these methods can account for feature interactions,
their main drawback is computational cost: a new model must be trained
for each candidate subset. This makes wrapper methods impractical for
high-dimensional turbulence datasets~\cite{chandrashekar2014, li2017}.

Another family of methods, known as embedded methods, alleviate this
computational cost by integrating feature selection into the training
process. A well-known example is the Least Absolute Shrinkage and
Selection Operator (LASSO)~\cite{tibshirani1996}, which introduces
regularization to enforce sparsity in regression coefficients. Related
approaches, such as the Elastic Net proposed by~\citet{zou2005},
extend LASSO by combining $\ell_1$ and $\ell_2$ penalties, offering
greater flexibility when dealing with correlated variables. In this
way, the optimization simultaneously minimizes prediction error while
pruning irrelevant features.

Despite their usefulness, the strategies above share a fundamental
limitation: they identify variables associated with the target, but
not necessarily those that drive its future
evolution~\cite{yu2020}. As a result, their predictions may fail to
generalize beyond the conditions observed during training. This
limitation has motivated the adoption of causality-based
approaches~\cite[e.g.][]{spirtes2001, lozano2022}, which aim to
recover the minimal set of causal parents of a target variable. By
identifying causal mechanisms rather than mere associations, this
family of methods promises more interpretable and robust
models. However, most existing algorithms still treat variables
individually and fail to capture the synergistic and redundant
interactions that characterize turbulence.

In this work, we introduce a method that directly addresses this gap
by grounding model input selection in causality while explicitly
accounting for multivariate interactions. Specifically, we employ the
Synergistic–Unique–Redundant Decomposition (SURD) of
causality~\cite{NatCom2024}, which disentangles the contribution of
each input feature into redundant, unique, and synergistic components
with respect to forecasting a target quantity. This cause-and-effect
perspective offers a principled approach for identifying the most
informative inputs and establishes fundamental limits on the
predictive capability of any forecasting model constructed from them.

The main contributions of the paper are:
\begin{enumerate}
\item Introducing a causality-driven approach for input selection in
  forecasting modeling of turbulence based on the SURD decomposition.
\item Demonstrating how unique, redundant, and synergistic causalities
  inform the construction of interpretable and parsimonious
  forecasting models.
\item Applying the methodology to turbulent channel-flow data to show
  that causal analysis identifies the set of input flow variables with
  superior predictive value.
\end{enumerate}

The remainder of the paper is organized as
follows. Section~\ref{sec:method} presents the methodology, including
the use of variational mutual information estimators. In
Section~\ref{sec:validation}, the approach is validated on a set of
illustrative examples. Section~\ref{sec:results} applies the approach
to turbulent channel flow and analyzes the causal structure of various
flow components. Finally, Section~\ref{sec:discussion} discusses the
broader implications for turbulence modeling, summarizes the main
findings, and outlines directions for future research.

\section{Methodology}
\label{sec:method}

Consider the collection of $N$ input variables evolving in space and
time given by the vector $\bQ = [Q_1(\bs{x}, t),Q_2(\bs{x}, t),$
  $\dots,Q_N(\bs{x}, t)]$. For example, $Q_i$ may represent the time
evolution of the streamwise velocity at a given distance from the
wall. The components of $\bQ$ are the input variables and are treated
as random variables. Our objective is to construct a forecasting model of the future of an output variable $Q_O^+$,
denoted by $Q_O^+ = Q_O(\bs{x}, t+\Delta T)$, where $\Delta T>0$ is an arbitrary time increment. To that end, we quantify the causal influence of input variables on the output and leverage this information to characterize the fundamental limits of predictability in forecasting models.

Our approach is structured in three main steps. First, we adopt the principle of \textit{forward-in-time propagation of information}---i.e., information flows only toward the future~\cite{lozano2022}---and quantify causality among variables in terms of information increments. We then decompose these causal influences into distinct interaction types: synergistic, unique, and redundant contributions. Second, we link these causal components to the \textit{information-theoretic irreducible error theorem}~\cite{lozano2022, yuan2024, yuan2025}, which enables us to quantify the minimum forecasting error achievable by any model, regardless of its form. Finally, we employ a mutual information neural estimator to compute causal relationships among high-dimensional variables, allowing the method to scale efficiently in complex systems.

\subsection{Observational causality with SURD}

For the first step, we adopt the definition of causality proposed in \citet{NatCom2024}, implemented through SURD. In this framework, causality is quantified as the increase in information about the future output $Q_O^+$ that is gained by observing individual or groups of past inputs $\bQ$. The information content in $Q_O^+$ is measured using Shannon entropy~\cite{shannon1948}, denoted as $H(Q_O^+)$, which reflects the average level of unpredictability---or expected surprise---associated with the outcomes of the random variable $Q_O^+$.

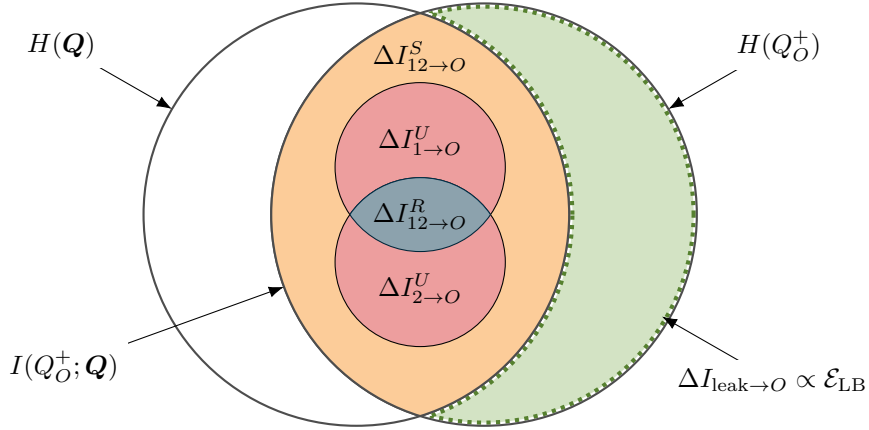
\begin{figure}[t!]
\vspace{1cm}
    \centering
    {\begin{tikzpicture}[scale=2.8,>={Latex[length=.2cm]}]
    \colorlet{c1}{myc1}
    \colorlet{c2}{myc2}
    \colorlet{ec}{black!25}
    \colorlet{ec}{myc4}

    \colorlet{MI}{myc3}
    \pgfmathsetmacro{\dis}{.3}
    \pgfmathsetmacro{\diss}{.4}
    \pgfmathsetmacro{\th}{.015}
    \begin{normalsize}

        \draw[ec!80!black,fill=ec!30,dotted,line width=0.55mm] (\dis,0) circle (1-\th);
        \begin{scope}
            \clip (-\dis,0) circle (1+\th);
            \draw[white,fill=white,line width=0.55mm] (\dis,0) circle (1-\th);
        \end{scope}
        \begin{scope}
            \clip (\dis,0) circle (1-\th);
            \draw[ec!80!black,fill=white,dotted,line width=0.55mm] (-\dis,0) circle (1+\th);
        \end{scope}

        \draw[black!70, thick] (-\dis,0) circle (1);
        \draw[black!70, thick] (\dis,0) circle (1);

        \begin{scope}
            \clip (-\dis,0) circle (1);
            \draw[draw=black!70,fill=MI!40, thick] (\dis,0) circle (1);
        \end{scope}
        \begin{scope}
            \clip (\dis,0) circle (1);
            \draw[black!70, thick] (-\dis,0) circle (1);
        \end{scope}

        \draw[fill=c2!45] (0,0.225) circle (\diss);
        \draw[fill=c2!45] (0,-0.225) circle (\diss);

        \begin{scope}
            \clip (0,0.225) circle (\diss);
            \draw[draw=c1,fill=c1!45] (0,-0.225) circle (\diss);
        \end{scope}
        \begin{scope}
            \clip (0,-0.225) circle (\diss);
            \draw[c1!80!black] (0,0.225) circle (\diss);
        \end{scope}

        \node              at (0,0.75) {$\Delta I _ {12\to O} ^ S$};
        \node              at (0,\diss-0.05) {$\Delta I _ {1\to O} ^ U$};
        \node              at (0,-\diss+0.05) {$\Delta I _ {2\to O} ^ U$};
        \node              at (0,0) {$\Delta I _ {12\to O} ^ R$};


        \draw[<-] (+\dis,0) ++ (30:1)  --+ (30:.6)  node[fill=white,pos=1] {$H(Q_O^+)$};
        \draw[<-] (-\dis,0) ++ (150:1) --+ (150:.6) node[fill=white,pos=1] {$H(\bQ)$};

        \draw[<-] (+\dis,0) ++ (200:1) --+ (200:1.1) node[fill=white,pos=1] {$I(Q_O^+;\bQ)$};
        \draw[<-] (+\dis,0) ++ (-30:0.965)  --+ (-30:.6)  node[fill=white,pos=1] {$\Delta I _{{\rm{leak}}\to O}\propto \mathcal{E}_{\rm{LB}}$};

    \end{normalsize}
\end{tikzpicture}}
    \caption{{SURD: Synergistic–Unique–Redundant Decomposition of causality.}
    Diagram of the decomposition of causal dependencies between the past variables 
    $\bQ = [Q_1, Q_2]$ and a future target $Q_O^+$ into their synergistic (S), unique (U) and redundant (R) components
      (in yellow, red, and blue, respectively). These contributions sum to the total mutual information 
    $I(Q_O^+; Q_1, Q_2)$, and relate to the Shannon entropies of the output 
    $H(Q_O^+)$ and the inputs $H(\bQ)$. The causality leak 
    $\Delta I_{\mathrm{leak}\to O}$ is highlighted in green and is approximately 
    proportional to the information-theoretic irreducible error 
    $\mathcal{E}_{\mathrm{LB}}$.}
    \label{fig:surd}
\end{figure}

Next, we decompose the information in $H(Q_O^+)$ into a
sum of information increments contributed by distinct types of
interactions from $\bQ$---namely, redundant, unique, and synergistic
components---using the principle of forward-in-time propagation of information~\cite{NatCom2024}:
\begin{equation}
\label{eq:surd}
    H(Q_O^+) = \sum_{\bi \in \mathcal{C}} \Delta I ^ R _ {\bi\to O} + \sum_{i=1}^N \Delta I ^ U _ {i\to O} + \sum_{\bi \in \mathcal{C}} \Delta I ^ S _ {\bi\to O} + \Delta I_{\rm{leak}\to O},
\end{equation}
where the terms $\Delta I^R_{\bi
  \rightarrow O}$, $\Delta I^U_{i \rightarrow O}$, and $\Delta
I^S_{\bi \rightarrow O}$ denote redundant, unique, and synergistic
causalities, respectively, from $\bQ$ to $Q_O^+$, and $\Delta I_{\rm{leak}\to O}$ is the causality from unobserved variables, referred to as the causality leak. Unique causalities
are associated with individual components of $\bQ$, while redundant
and synergistic causalities emerge from interactions among groups of
variables. The set $\mathcal{C}$ includes all subsets of indices with
cardinality greater than one, i.e., $\mathcal{C} = \{ \bi \subseteq
\{1, \dots, N\} \mid |\bi| > 1 \}$. For instance, for $N=2$, Eq. \ref{eq:surd} reduces to $H(Q_O^+) = \Delta I ^ R _ {12\to O} + \Delta I ^ U _ {1\to O} + \Delta I ^ U _ {2\to O} + \Delta I ^ S _ {12\to O} + \Delta I_{\rm{leak}\to O}$. Figure \ref{fig:surd} shows the diagram of the redundant, unique, and synergistic causalities for $N = 2$.

To quantify the causal components in Eq. \ref{eq:surd}, we rely on the concept of mutual
information \cite{shannon1948} between the target variable $Q_O^+$ and combinations of the input variables $\bQ_\bi$. This quantity can be
mathematically described as:
\begin{equation} \label{eq:mi}
I(Q_O^+; \bQ_\bi) = \int_{q_O^+\in Q_O} \int_{\bq_\bi\in\bQ_\bi} p(q_O^+, \bq_\bi) \log_2 \left( \frac{p(q_O^+, \bq_\bi)}{p(q_O^+) p(\bq_\bi)} \right) {\rm{d}}\bq_\bi\,{\rm{d}}q_O^+,
\end{equation}
where $p(q_O^+, \bq_\bi)$, $p(q_O^+)$, and $p(\bq_\bi)$ denote the joint and
marginal probability density functions of the output and input
variables, respectively, and $q_O^+$ and $\bq_\bi$ represent particular
values of the output and input variables. Mutual information measures
how different the joint probability density function $p(q_O^+, \bq_\bi)$
is from the hypothetical distribution $p(q_O^+)p(\bq_\bi)$, where $q_O^+$
and $\bq_\bi$ are assumed to be independent. For instance, if $Q_O^+$ and
$\bQ_\bi$ are not independent, then $p(q_O^+, \bq_\bi)$ will differ
significantly from $p(q_O^+)p(\bq_\bi)$. Hence, we assess causality by
examining how the probability of $Q_O^+$ changes when accounting for
$\bQ_\bi$.

Then, we quantify the information increments $\Delta I$ about $Q_O^+$ obtained by observing individual components or groups of components from $\bQ$. This procedure enables the decomposition of the mutual information $I(Q_O^+;\bQ)$ into redundant, unique, and synergistic contributions. For the case $N=2$, Figure~\ref{fig:surd} illustrates the decomposition: $I(Q_O^+;\bQ) = \Delta I^{R}_{12\to O} + \Delta I^{U}_{1\to O} + \Delta I^{U}_{2\to O} + \Delta I^{S}_{12\to O}$.
The mathematical definitions of these terms are provided in \S\ref{sec:app}; here, we focus on their interpretation:
\begin{itemize}
    \item \textit{Redundant causality} from a subset $\bQ_{\bi} = \{
      Q_{i_1}, Q_{i_2}, \dots \} \subseteq \bQ$ to $Q_O^+$, denoted by
      $\Delta I^R_{\bi \rightarrow O}$, is the information about the
      output that is identically present in all variables within the
      group $\bQ_{\bi}$. Redundant causality arises when each variable
      in the group individually contains the same information about
      the target. 
    
    \item \textit{Unique causality} from an individual variable $Q_i$
      to $Q_O^+$, denoted by $\Delta I^U_{i \rightarrow O}$, is the
      information about the output that is available exclusively
      through $Q_i$ and cannot be recovered from any other single
      variable. Unique causality indicates that $Q_i$ provides
      critical information not found elsewhere in the set of
      individual variables. 
    
    \item \textit{Synergistic causality} from a subset $\bQ_{\bi} = \{
      Q_{i_1}, Q_{i_2}, \dots \} \subseteq \bQ$ to $Q_O^+$, denoted by
      $\Delta I^S_{\bi \rightarrow O}$, corresponds to the information
      that can only be accessed when all variables in the group are
      considered jointly. Synergy captures higher-order interactions,
      where the collective observation of variables reveals
      information that is absent when they are observed
      individually. 
\end{itemize}

\subsection{Causality-driven irreducible model error}

In the second step, we relate the redundant, unique, and synergistic causalities to the forecasting error of a model. To this end, we build upon the information-theoretic irreducible error theorem introduced by \citet{yuan2025}. The theorem establishes that the minimum forecasting error achievable by any model, denoted by $\mathcal{E}_{\rm LB}$, corresponds to the uncertainty that remains in the output $Q_O^+$ after observing the inputs $\bQ$. This residual uncertainty, quantified by the conditional entropy $H(Q_O^+ \mid \bQ)$, matches the concept of causality leak as defined in Eq.~\ref{eq:error}. This connection allows us to attribute the contributions of each causal component---redundant, unique, and synergistic---to the lower bound $\mathcal{E}_{\rm LB}$.

In particular, let $\mathcal{F}$ denote the space of all possible forecasting models of $Q_O^+$ that take $\bQ$ as input. For any model $f \in \mathcal{F}$, producing the prediction $\hat{Q}_O = f(\bQ)$, the expected error under an $L_p$-norm is bounded as:

%
\begin{equation}
\label{eq:error}
    \min_{f \in \mathcal{F}} \| Q_O^+ - \hat{Q}_O^+ \|_p \geq 
    {\prod_{\bi \in \mathcal{C}} e^{-\Delta I^R_{\bi \rightarrow O}}} \cdot
    {\prod_{i=1}^N e^{-\Delta I^U_{i \rightarrow O}}} \cdot 
    {\prod_{\bi \in \mathcal{C}}e^{-\Delta I^S_{\bi \rightarrow O}}}
    \cdot c\big[p, H(Q_O^+)\big] \equiv \mathcal{E}_{\rm{LB}},
\end{equation}
where the function $c\big[p, H(Q_O^+)\big]$ depends only on the choice
of norm $p$ and the differential entropy of the output variable
$H(Q_O^+)$. The general proof for this bound and the explicit form of
the constant $c\big[p, H(Q_O^+)\big]$ for the Rényi entropy of order
$\alpha$ is given in \citet{yuan2025}. 
%
The terms $e^{-\Delta I^R_{\bi \rightarrow O}}$, $e^{-\Delta I^U_{i \rightarrow O}}$, and $e^{-\Delta I^S_{\bi \rightarrow O}}$ denote the contributions of the redundant, unique, and synergistic causal components to the minimum forecasting error, respectively. Here, we focus on the connection between each of the causal components $\Delta I$ and the construction of forecasting models:
\begin{itemize}
    \item \textit{Redundant error contributions} from a subset $\bQ_{\bi} = \{
      Q_{i_1}, Q_{i_2}, \dots \} \subseteq \bQ$ to $Q_O^+$, denoted by
     $e^{-\Delta I^R_{\bi \rightarrow O}}$, represent the contributions to the error bound of the redundant causality from the group $\bQ_{\bi}$ about $Q_O^+$. 
      In this case, forecasting models for $Q_O^+$ can be
      simplified by selecting the most convenient variable from the
      redundant set and disregarding the rest.
    
    \item \textit{Unique error contributions} from an individual variable $Q_i$
      to $Q_O^+$, denoted by $e^{-\Delta I^U_{i \rightarrow O}}$, represent the contributions to the error bound of the unique causality from $Q_i$ about $Q_O^+$. 
      Therefore, forecasting models for $Q_O^+$
      should always retain $Q_i$ as input, since its information
      cannot be found in any other variable alone.
    
    \item \textit{Synergistic error contributions} from a subset $\bQ_{\bi} = \{
      Q_{i_1}, Q_{i_2}, \dots \} \subseteq \bQ$ to $Q_O^+$, denoted by
      $e^{-\Delta I^S_{\bi \rightarrow O}}$, correspond to the contributions to the error bound of the synergistic causality from the group $\bQ_\bi$ about $Q_O^+$.
      Therefore, it is crucial for models to incorporate
      all variables in $\bQ_{\bi}$ as inputs to ensure accurate
      forecasts.
\end{itemize}

Figure \ref{fig:surd} shows the relationship between the redundant, unique, and
synergistic causalities with the
output information $H(Q_O^+)$ and the minimum forecast error
$\mathcal{E}_{\rm{LB}}$. In this case, the expected error in Eq.
\ref{eq:error} is bounded as:
\begin{equation}
\label{eq:error_2vars}
    \mathcal{E}_{\rm{LB}} =  
    e^{-\Delta I^R_{12 \rightarrow O}} \cdot e^{-\Delta I^U_{1 \rightarrow O}}
    \cdot e^{-\Delta I^U_{2 \rightarrow O}}
    \cdot e^{-\Delta I^S_{12 \rightarrow O}} \cdot
    c\big[p, H(Q_O^+)\big].
\end{equation}

The diagram in Figure \ref{fig:surd} implies that a perfect
prediction is achievable only when the inputs $\bQ$ fully determine
the output $Q_O^+$, i.e., $Q_O^+ = \hat{Q}_O^+$. In the continuous
case, this condition corresponds to any of the causal terms $\Delta I$ diverging to infinity, leading the irreducible error bound in
Eq.~\ref{eq:error} to vanish asymptotically as $e^{-\infty}\to
0$. Conversely, when some of the information required to predict
$Q_O^+$ is absent from $\bQ$, the causal terms $\Delta I$ remain finite,
and the irreducible error remains strictly positive. This lower bound
cannot be reduced by increasing model complexity, as it reflects a
fundamental information-theoretic limit imposed by the incompleteness
of the input.

\subsection{Mutual information estimation in high-dimensional spaces}
\label{sec:mine}

Evaluating the causal contributions discussed above requires computing the mutual
information between the set of input variables $\bQ_{\bi}$ and the
output variable $Q_O^+$. However, this task becomes particularly
challenging in high-dimensional settings, such as those encountered in
turbulent flows. The main difficulty arises from the intractability of
accurately estimating the joint and marginal probability distributions
$p(q_O^+, \bq_{\bi})$, $p(q_O^+)$, and $p(\bq_{\bi})$ when both
$q_O^+$ and $\bq_{\bi}$ lie in high-dimensional spaces.

To illustrate this, consider a case where $Q_O^+$ represents a
two-dimensional field of wall-shear stress in a turbulent channel
flow, and $\bQ_{\bi}$ corresponds to the streamwise velocity field at
a given wall-normal location. Suppose a naive binning approach is used
to estimate probabilities, where both fields are discretized over a $5
\times 5$ grid and their joint distribution is computed using a
histogram-based method with 10 bins per variable. The resulting joint
space would contain approximately $10^{50}$ bins, requiring at least
an order of magnitude more independent samples to obtain statistically
meaningful estimates---a clearly infeasible demand.

To overcome this challenge, we employ a variational formulation of
mutual information known as the \textit{Donsker–Varadhan (DV)
  representation}. This representation expresses the mutual
information as a functional optimization problem over a class of
real-valued functions $g \in \mathcal{G}$, which can be parametrized
and optimized directly from data:
\begin{equation} \label{eq:dv}
I(Q_O^+; \bQ_{\bi}) \geq \sup_{g \in \mathcal{G}} \left( \mathbb{E}_{p(q_O^+, \bq_{\bi})} \left[ g(q_O^+, \bq_{\bi}) \right] - \log \, \mathbb{E}_{p(q_O^+)p(\bq_{\bi})} \left[ e^{g(q_O^+, \bq_{\bi})} \right] \right),
\end{equation}
where $\mathbb{E}_{p(\bq_{\bi})}\left[ \cdot \right]$ denotes the
expectation operator under the distribution $p(\bq_{\bi})$ and
similarly for other terms. This bound consists of two expectations:
\begin{itemize}
\item The first term $\mathbb{E}_{p(q_O^+, \bq_{\bi})} \left[ g(q_O^+, \bq_{\bi})
  \right]$ depends on samples drawn from the {joint distribution}
  $p(q_O^+, \bq_{\bi})$. It rewards the function for assigning high
  importance to true input–output pairs.
\item The second term $\log \, \mathbb{E}_{p(q_O^+)p(\bq_{\bi})}
  \left[ e^{g(q_O^+, \bq_{\bi})} \right]$ is computed over the
       {product of marginals} and penalizes functions that also assign
       high importance to independent input–output combinations.
\end{itemize}
The optimal function $g^*$ that achieves equality in this bound is the
log-density ratio $\log \frac{p(q_O^+, \bq)}{p(q_O^+)p(\bq)}$, which
directly characterizes the mutual dependence between $q_O^+$ and
$\bq_{\bi}$, as shown in Eq. \ref{eq:mi}. In practice, the closer the
learned function $g$ approximates this optimal log-ratio, the tighter
the bound becomes, which enables the estimation of the mutual
information without the need for explicit density modeling.

\begin{figure}[t!]
    \centering

    \begin{tikzpicture}

        \node[anchor=south west, inner sep=0] (img) at (0,0) {
            \includegraphics[width=\linewidth, trim=2.75cm 30pt 7.6cm 10pt, clip]{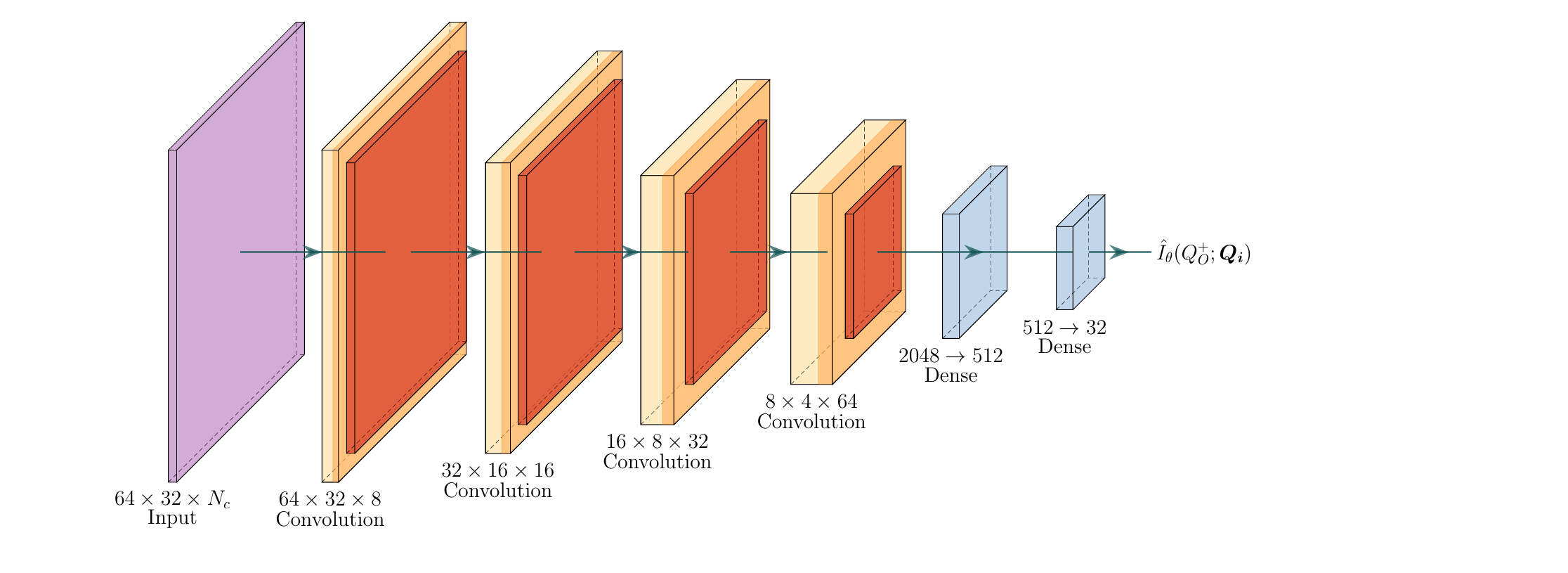}
        };


    \end{tikzpicture}

    \caption{Schematic of the architecture for the mutual information
      estimator in Eq. (\ref{eq:mine}). The numbers below the
      convolution block denote the size of the filter and the number
      of channel applied at each layer. $N_c$ represents the number of
      channels of the input layer, which denotes the number of
      variables ($Q_O^+$ and $\bQ_{\bi}$) between which the mutual
      information is estimated.}
    \label{fig:mine-arch}
\end{figure}

Several practical estimators have been developed based on the
variational representation in Eq.~\ref{eq:dv}, including
MINE~\cite{belghazi2018}, InfoNCE~\cite{oord2019}, and
TUBA~\cite{poole2019}. These methods differ primarily in how the
variational function is parametrized and how the expectations over the
joint and marginal distributions are estimated in practice. In this
work, we adopt the MINE (Mutual Information Neural Estimation)
method~\cite{belghazi2018}, which directly implements the
Donsker--Varadhan bound using a neural network to approximate the
function $g$. Specifically, the function $g$ is parametrized by a
neural network $g_\theta$ with learnable parameters $\theta$, and the
mutual information is estimated as:
\begin{equation} \label{eq:mine}
\hat{I}_{\theta}(Q_O^+; \bQ_{\bi}) = \frac{1}{m} \sum_{k=1}^{m} g_{\theta}(q_O^{(k)}, \bq_{\bi}^{(k)}) - \log \left( \frac{1}{m} \sum_{k=1}^{m} e^{g_{\theta}(\tilde{q}_O^{(k)}, \bq^{(k)}_{\bi})} \right),
\end{equation}
where $\{(q_O^{(k)}, \bq^{(k)}_{\bi})\}_{k=1}^m$ are mini-batch
samples drawn from the joint distribution $p(q_O^+, \bq_{\bi})$, and
$\{(\tilde{q}_O^{(k)}, \bq^{(k)}_{\bi})\}_{k=1}^m$ are surrogate
samples used to approximate the product of marginals
$p(q_O^+)p(\bq_{\bi})$. The marginal samples are constructed by
independently shuffling the output values $\{q_O^{(k)}\}$ across the
batch while keeping the inputs $\{\bq^{(k)}_{\bi}\}$ fixed. This
permutation breaks any statistical dependence between $q_O^+$ and
$\bq_{\bi}$, thus providing samples that approximate the assumption of
independence under the marginal product distribution. The contrast
between the importance assigned to true (joint) and shuffled
(independent) pairs allows the estimator to learn a function
$g_\theta$ that approximates the log-density ratio.

The network $g_\theta$ is trained by maximizing $\hat{I}_\theta$ using
stochastic gradient ascent over minibatches.
Figure~\ref{fig:mine-arch} illustrates the architecture of the mutual
information estimator used in this work. The input layer receives the
spatial fields of the target variable $Q_O^+$ and the set of candidate
inputs $\bQ_{\bi}$, stacked along the channel dimension ($N_c$). These
inputs are processed through a sequence of convolutional layers, each
reducing the spatial resolution while increasing the number of feature
channels to progressively extract higher-level representations. The
final layers map these features to a scalar estimate of
$\hat{I}_\theta$.

\section{Validation}
\label{sec:validation}

We consider two benchmark cases to illustrate how SURD causalities can
guide the selection of input variables in forecasting models. Each
case is designed to exhibit a different type of collider effect, in
which two input variables, $Q_2$ and $Q_3$, collectively influence the
future state of the output variable $Q_1$. For simplicity, the
variables $Q_i$ are considered time-dependent only, although the
formulation introduced above is applicable to variables that are
functions of space and time.

\subsection{Collider with synergistic variables}

\begin{figure}[t!]
    \centering
    \scalebox{1.2}{
    \begin{minipage}{\textwidth}
    \centering
    \hspace{-2.75cm}
    \begin{minipage}{0.175\textwidth}
    \scalebox{0.95}{
        $\hspace{0.cm}{\hbox{


\begin{tikzpicture}[cir/.style={circle,draw=black!70!white,,thick,inner sep=.5em},
    >={Latex[length=.2cm]},scale=.8]
    

    \colorlet{cA}{Q1!30!white}
    \colorlet{cB}{Q2!30!white}
    \colorlet{cC}{Q3!30!white}

    \node [cir,fill=cB] (q1) at (.73,0)   {$Q_1$}; 
    \node [cir,fill=cA] (q2) at (-1,-1) {$Q_2$}; 
    \node [cir,fill=cC] (q3) at (-1,+1) {$Q_3$}; 
    
    \node [anchor=east] at (q1.west) {\scriptsize synergistic collider}; 

    \draw[decorate,decoration={snake,post length=4pt,amplitude=.2em}, thick,->]
    (-2.3,1) node[anchor=east] {$W_3$} -- (q3);
    \draw[decorate,decoration={snake,post length=4pt,amplitude=.2em}, thick,->]
    (-2.3,-1) node[anchor=east] {$W_2$} -- (q2);
    \draw[decorate,decoration={snake,post length=4pt,amplitude=.2em}, thick,->]
    (.73,-1.3) node[anchor=north] {$W_1$} -- (q1);

    \path[thick,->]
    (q3) edge[bend left] node [left] {} (q1)
    (q2) edge[bend right] node  [left] {} (q1);

    \path[thick,<-] (q2) edge[loop below] node {} (q2);
    \path[thick,<-] (q3) edge[loop above] node {} (q3);

\end{tikzpicture}

}}$
    }
    \end{minipage}
    \hspace{0.1\textwidth}
    \begin{minipage}{0.3395\textwidth}
    \begin{tikzpicture}[scale=1, transform shape]
      \node[inner sep=0pt, outer sep=0pt, anchor=south west] (img) at (0,0)
      {\includegraphics[width=\linewidth,trim={0 0 2.5cm 1.25cm},clip]{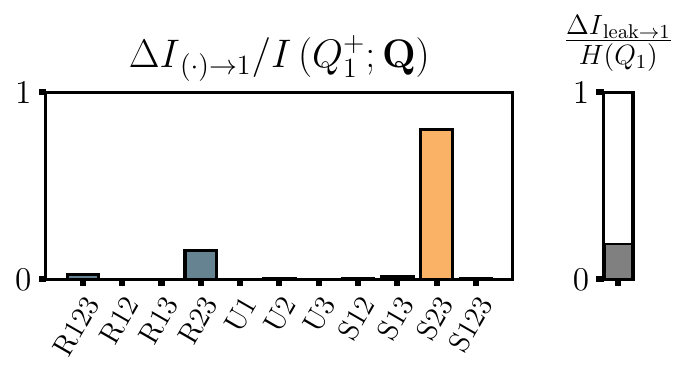}};    
      \node[above=-3pt of img.north, align=center, fill=white] {$\quad$SURD causalities to $Q_1^+$};
    \end{tikzpicture}
    \end{minipage}
    \end{minipage}}

    \vspace{0.1cm}
    \begin{minipage}{\textwidth}
    \includegraphics[height = 0.197\textwidth,trim={0 0 0 0},clip]{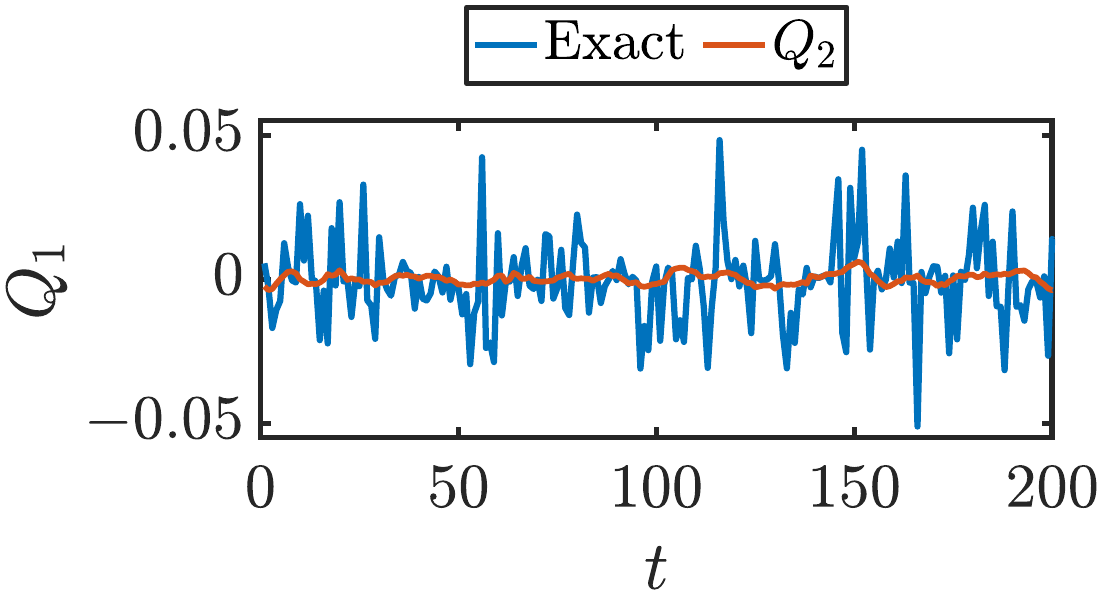}
    \hfill
    \includegraphics[height = 0.197\textwidth,trim={4.1cm 0 0 0},clip]{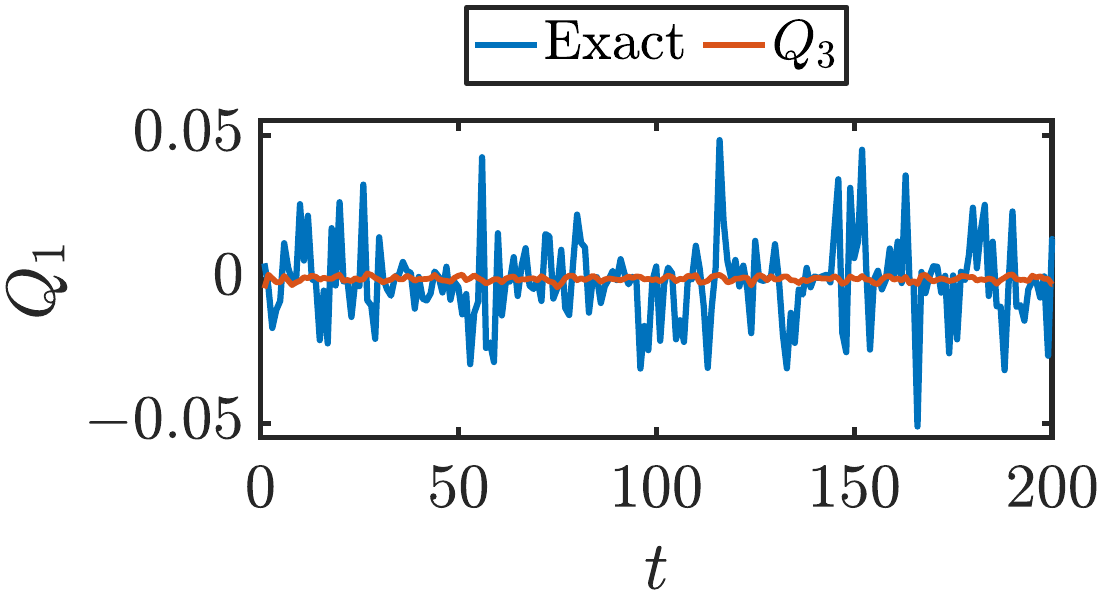}
    \hfill
    \includegraphics[height = 0.197\textwidth,trim={4.1cm 0 0 0},clip]{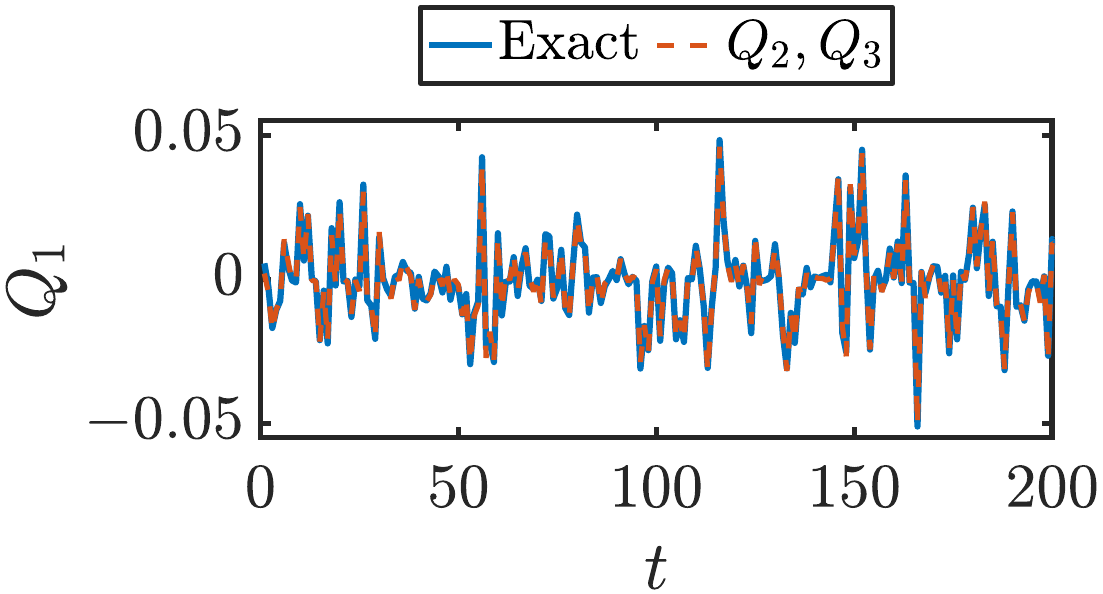}
    \end{minipage}

    \caption{{Benchmark case with synergistic collider variables where
        $Q_2$ and $Q_3$ collectively influence the future of
        $Q_1$}. (Top left panel) Schematic of the functional
      dependence among variables and system equations, where $W_i$
      represents unobserved, stochastic forcing on the variable
      $Q_i$. (Top right panel) Results from SURD with redundant (R),
      unique (U) and synergistic (S) causalities in blue, red and
      yellow, respectively. The notation employed is such that R123
      denotes $\Delta I_{123\rightarrow j}^R$ and so on. (Bottom
      panel) Comparative performance of LSTM models for forecasting
      the future of $Q_1$ using different input variables.  The legend
      indicates the variables used as input to the LSTM model along
      with the exact solution.}
    \label{fig:synergistic}
\end{figure}
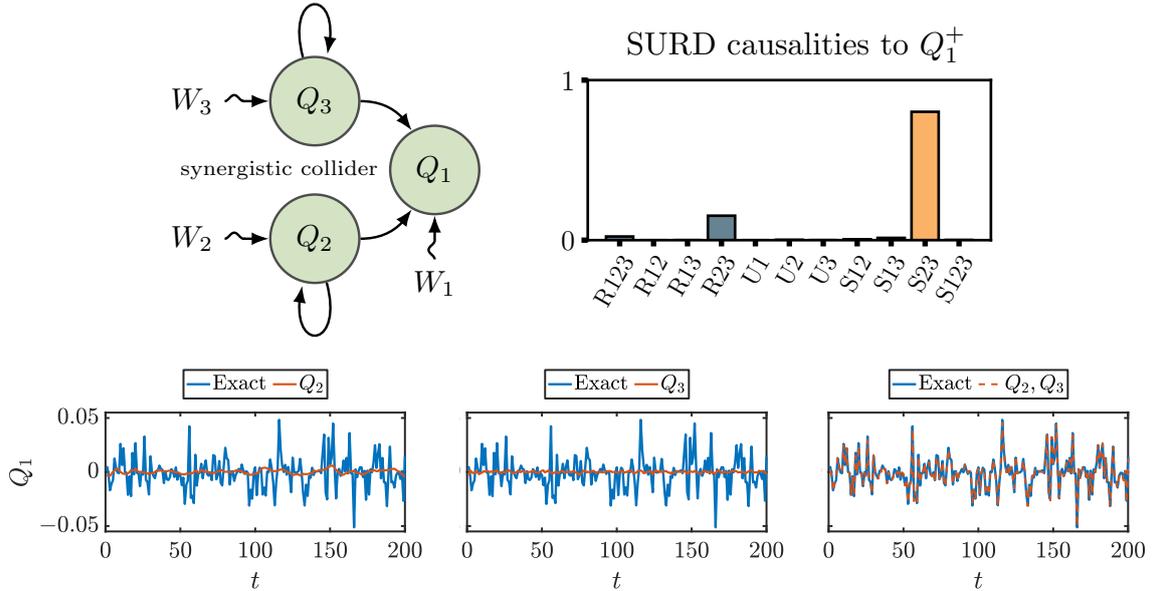

The first case investigated corresponds to a collider where the pair
$[Q_2,Q_3]$ influences $Q_1^+$ synergistically, i.e., the predictive
information about $Q_1^+$ arises when the two inputs are considered
together rather than individually. This implies that $Q_2$ and $Q_3$
behave as a single random variable that drives $Q_1$. The system is
defined by the following stochastic recurrence relations:
\begin{flalign}
Q_1({n+1}) &= \sin\left[Q_2(n)Q_3(n)\right] + 0.001W_1(n)\\
Q_2(n+1) &= 0.5Q_2(n) + 0.1W_2(n)\\
Q_3({n+1}) &= 0.5Q_3(n) + 0.1W_3(n),
\end{flalign}
where $W_i$ represents unobserved, stochastic forcing on the variable
$Q_i$ and $n$ indicates the discrete time step. Figure
\ref{fig:synergistic} shows a diagram with the relationships among the
variables, along with the results derived from SURD for the output
variable $Q_1^+ \equiv Q_1(n+1)$. The notation employed for SURD
causalities is such that R23 denotes $\Delta I _ {23\to 1} ^ R$, and
so on. The results reveal that the dominant causal contribution is the
synergistic causality from $Q_2$ and $Q_3$ to $Q_1^+$, quantified by
$\Delta I^{S}_{23 \to 1}$. This term accounts for approximately 80\%
of the total SURD causalities to $Q_1^+$. This indicates that the
minimum forecasting error, $\elb$, is achieved only when both
variables are considered jointly, while the reduction in error
attainable using $Q_2$ or $Q_3$ alone is negligible. Consequently, an
effective forecasting model for $Q_1^+$ must incorporate both $Q_2$
and $Q_3$ simultaneously in order to reach the theoretical limit of
predictive accuracy.

To test these insights in practice, we construct a set of forecasting
models based on long-short-term memory (LSTM) artificial neural
networks trained to predict $Q_1(n+1)$, using the exact values of
$Q_1(n)$, $Q_2(n)$, and $Q_3(n)$. Several models are trained using
different sets of input variables. The network architecture includes a
sequence input layer with the corresponding number of input features,
an LSTM layer with 200 hidden units to capture temporal dependencies
between the signals, and a fully connected layer to map the previous
layer to the output variable. The network is trained using an Adam
optimizer with a maximum of 200 epochs and an initial learning rate of
0.01, which is reduced by a factor of 0.3 with a period of 125
iterations.

The results for the predictions of the forecasting models are shown in
Figure \ref{fig:synergistic}, where we can clearly observe that the
forecasting performance of the models using $[Q_2, Q_3]$ significantly
surpasses those that include either variable alone. This outcome is
consistent with the synergistic causality detected by SURD, where
$Q_2$ and $Q_3$ collectively drive the future of $Q_1$. Generally,
this confirms that accurate forecasting of variables affected by
synergistic causalities is achievable only when all synergistically
interacting variables are incorporated into the model.

\begin{figure}[t!]
    \centering
    \scalebox{1.2}{
    \begin{minipage}{\textwidth}
    \centering
    \hspace{-2.75cm}
    \begin{minipage}{0.175\textwidth}
    \scalebox{0.95}{
        $\hspace{0.cm}{\hbox{

\newcommand{\triple}[3]{
  \draw[thick] (#1) -- (#2);
  \draw[thick] ([xshift=0.7mm]#1.south) -- ([xshift=0.7mm]#2.north);
  \draw[thick] ([xshift=-0.7mm]#1.south) -- ([xshift=-0.7mm]#2.north);
}

\begin{tikzpicture}[cir/.style={circle,draw=black!70!white,,thick,inner sep=.5em},
    >={Latex[length=.2cm]},scale=.8]
    

    \colorlet{cA}{Q1!30!white}
    \colorlet{cB}{Q2!30!white}
    \colorlet{cC}{Q3!30!white}

    \node [cir,fill=cB] (q1) at (1.73,0)   {$Q_1$}; 
    \node [cir,fill=cA] (q2) at (0,-1)     {$Q_2$}; 
    \node [cir,fill=cC] (q3) at (0,+1)     {$Q_3$}; 

    \node[anchor=south] at (q3.north) {\scriptsize redundant collider};
    \draw[decorate,decoration={snake,post length=4pt,amplitude=.2em}, thick,->]
    (-1.3,-1) node[anchor=east] {$W_2$} -- (q2);
    \draw[decorate,decoration={snake,post length=4pt,amplitude=.2em}, thick,->]
    (1.73,-1.3) node[anchor=north] {$W_1$} -- (q1);

    \path[thick,->]
    (q3) edge[bend left] node [left] {} (q1)
    (q2) edge[bend right] node  [left] {} (q1);

    \path[thick,<-] (q2) edge[loop below] node {} (q2);
    \path[thick,<-] (q1) edge[loop right] node {} (q1);
    
    \triple{q3}{q2};
\end{tikzpicture}

}}$
        }
    \end{minipage}
    \hspace{0.1\textwidth}
    \begin{minipage}{0.3395\textwidth}
    \vspace{-0.325cm}
    \begin{tikzpicture}[scale=1, transform shape]
      \node[inner sep=0pt, outer sep=0pt, anchor=south west] (img) at (0,0)
      {\includegraphics[width=\linewidth,trim={0 0 2.5cm 1.25cm},clip]{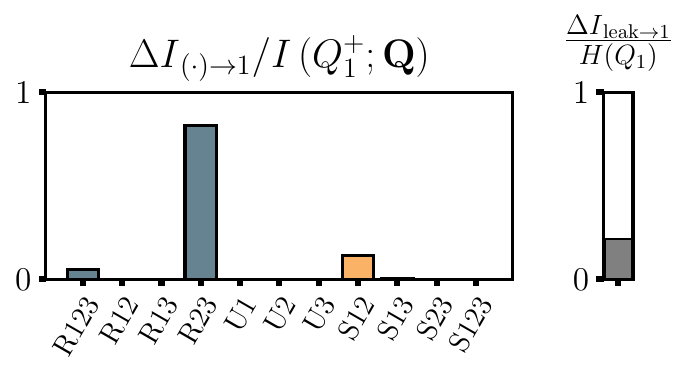}};    
      \node[above=-3pt of img.north, align=center, fill=white] {$\quad$SURD causalities to $Q_1^+$};
    \end{tikzpicture}
    \end{minipage}
    \end{minipage}}

    \vspace{0.1cm}
    \begin{minipage}{\textwidth}
    \includegraphics[height = 0.19\textwidth,trim={0 0 0 0},clip]{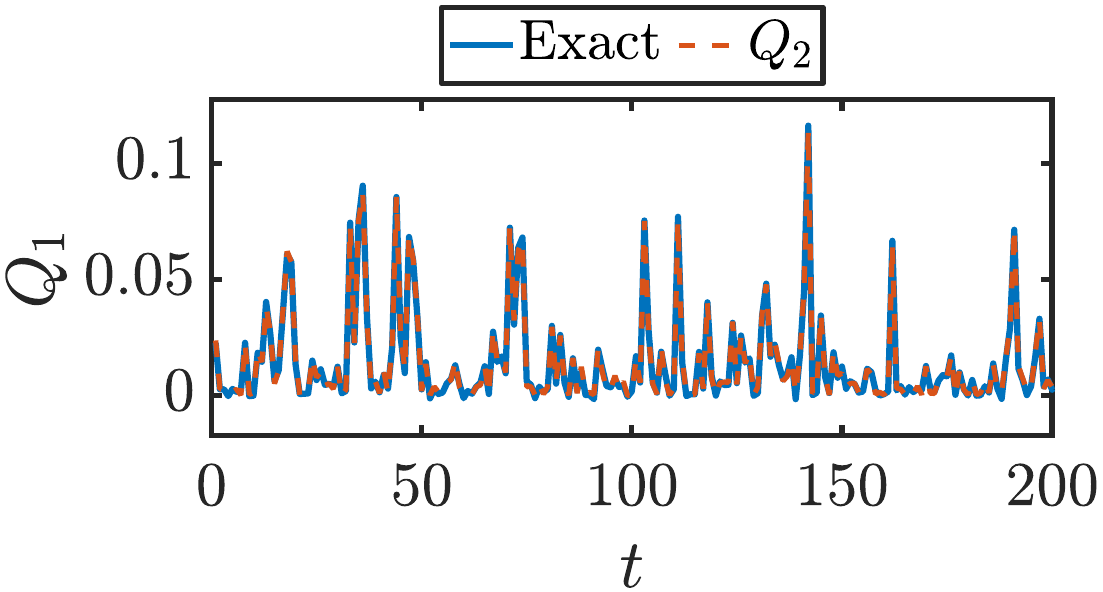}
    \hfill
    \includegraphics[height = 0.19\textwidth,trim={3.25cm 0 0 0},clip]{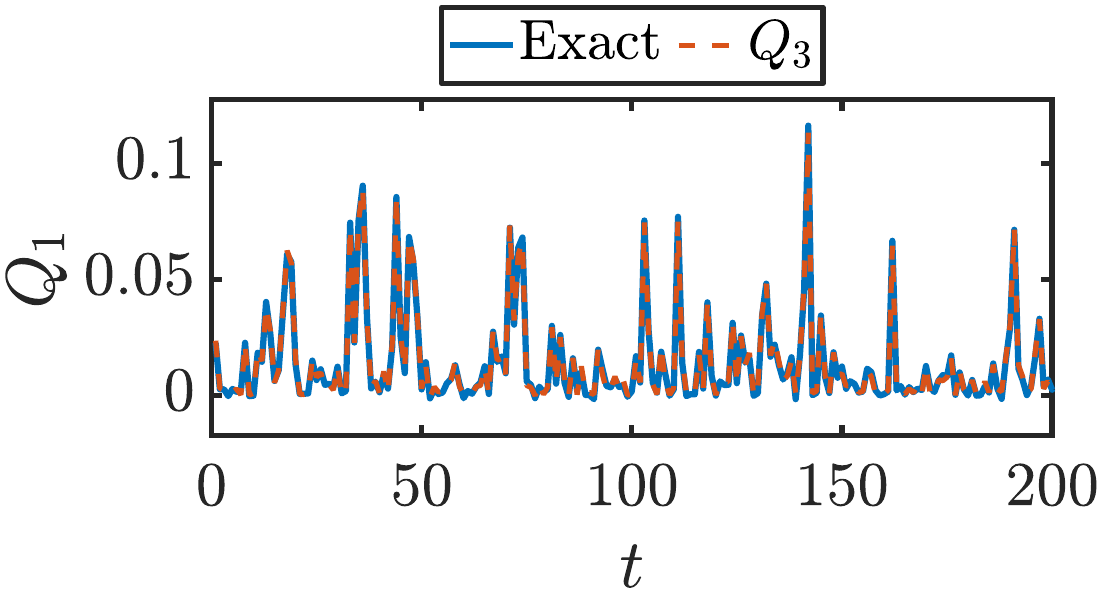}
    \hfill
    \includegraphics[height = 0.19\textwidth,trim={3.25cm 0 0 0},clip]{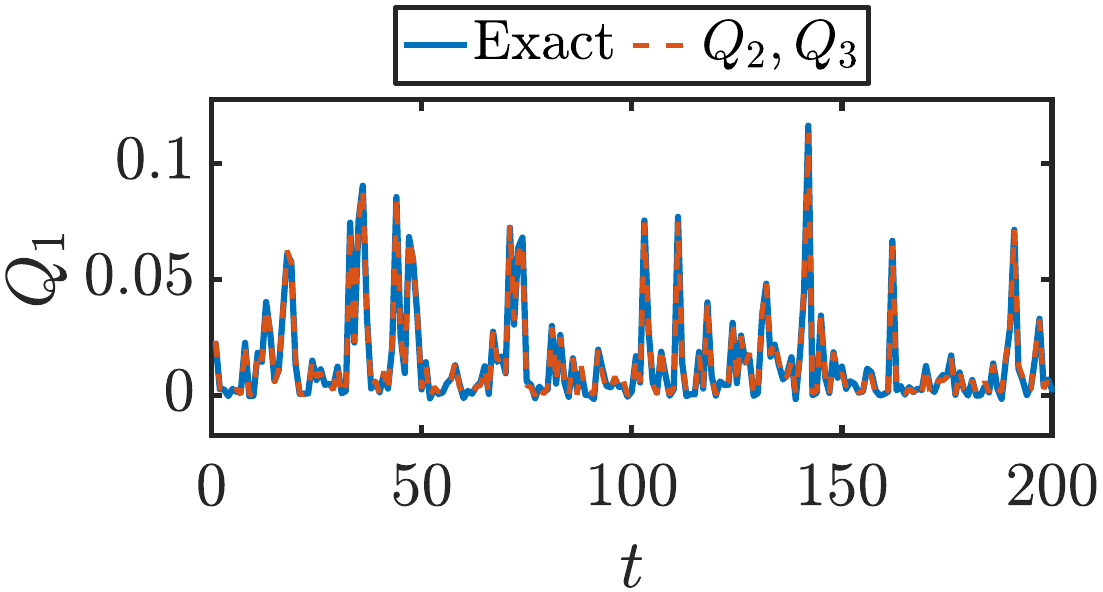}
    \end{minipage}
    \caption{{Benchmark case with redundant collider variables where
        the duplicated variables $Q_2$ and $Q_3$ collectively
        influence the future of $Q_1$}. (Top left panel) Schematic of
      the functional dependence among variables and system equations,
      where $W_i$ represents unobserved, stochastic forcing on the
      variable $Q_i$. The symbol $\equiv$ indicates that variables
      $Q_2$ and $Q_3$ are identical.  (Top right panel) Results from
      SURD with redundant (R), unique (U) and synergistic (S)
      causalities in blue, red and yellow, respectively. The notation
      employed is such that R123 denotes $\Delta I_{123\rightarrow
        j}^R$ and so on. (Bottom panel) Comparative performance of
      LSTM models for forecasting the future of $Q_1$ using different
      input variables.  The legend indicates the variables used as
      input to the LSTM model along with the exact solution.}
    \label{fig:redundant}
\end{figure}
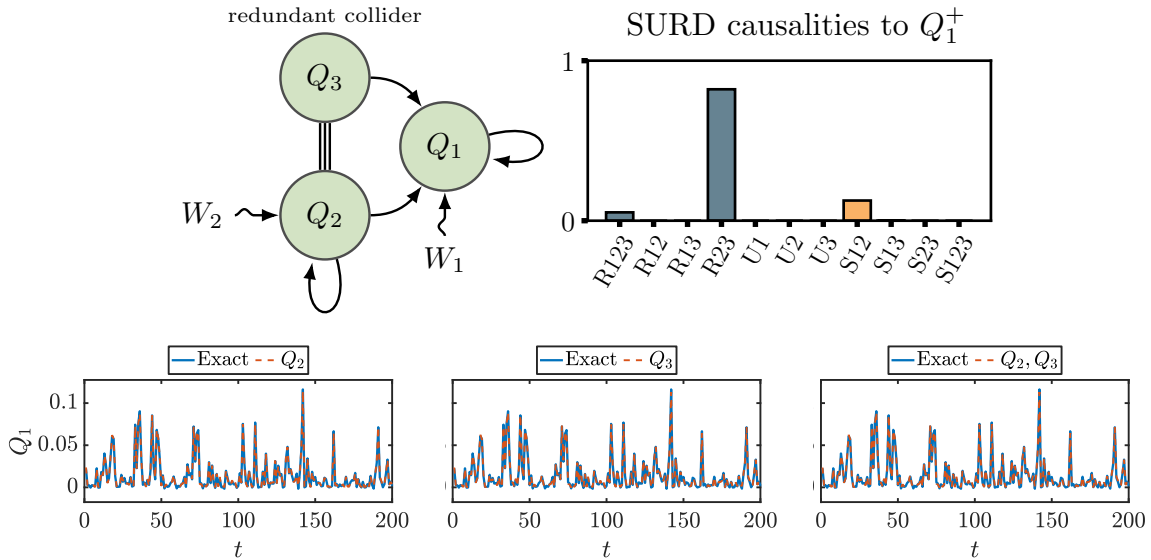

\subsection{Collider with redundant variables}

The second case explores the fundamental interaction $Q_2\equiv Q_3\to Q_1$, where 
$Q_3$ is identical to $Q_2$. In this scenario, $Q_2$ and $Q_3$ equally influence the 
future outcomes of $Q_1$. The governing equations of the system are:
\begin{flalign*}
Q_1({n+1}) &= 0.1Q(n) + \sin\left[Q_2(n)Q_3(n)\right] + 0.001W_1(n)\\
Q_2({n+1}) &= 0.5Q_2(n) + 0.1W_2(n)\\
Q_3({n+1}) &\equiv Q_2(n+1).
\end{flalign*}

The results shown in Figure~\ref{fig:redundant} indicate that SURD
identifies $\Delta I^{R}_{23 \to 1}$ as the dominant causal
contribution to $Q_1^+$. This redundant term accounts for 87\% of the
total causality, highlighting the duplicated influence of $Q_2$ and
$Q_3$ on the future state of $Q_1^+$. The fact that redundancy
dominates the causal structure implies that either $Q_2$ or $Q_3$ is
equally useful for predicting the target. Consequently, an accurate
forecasting model need only include one of them, as each alone
provides access to the redundant information critical for predicting
$Q_1^+$.

Figure~\ref{fig:redundant} also shows the predictions of forecasting
models trained with different input variables, obtained using an LSTM
network analogous to that employed in the previous system. We can
observe that the predictive accuracy of the forecasting model is not
compromised by using either $Q_2$ or $Q_3$. Furthermore, when both
variables are used simultaneously, the forecasting accuracy is neither
compromised nor improved.  Hence, in scenarios characterized by high
redundancy, compact predictive models can be optimized by selecting
the most convenient variable from the redundant set. This
interchangeability provides a strategic advantage in model
construction, allowing for the selection of variables based on
practical considerations, such as measurement ease or data
availability.

\section{Results}
\label{sec:results}

In this section, we investigate the causal relationship between the
wall-shear stress (output) and velocity fluctuations (input) in a
turbulent channel flow, using data from a direct numerical simulation
(DNS) at a friction Reynolds number $Re_\tau = u_\tau h / \nu \approx
180$, where $u_\tau$ is the friction velocity, $h$ is the channel
half-height, and $\nu$ is the kinematic viscosity. The computational
domain spans $L_x \times L_y \times L_z = \pi h \times 2h \times
\frac{\pi}{2} h$, with periodic boundary conditions in the streamwise
($x$) and spanwise ($z$) directions, and no-slip conditions at the
walls ($y = 0$ and $y = 2h$). The simulation is driven by a constant
streamwise mass flux and fully resolves all spatial and temporal
turbulence scales. Details of the numerical solver and simulation
setup can be found in \cite{lozano2020}.

The resulting database contains approximately $7\times10^5$
time-resolved snapshots of the three velocity fluctuation components:
streamwise $u(\bs{x}, t)$, wall-normal $v(\bs{x}, t)$, and spanwise
$w(\bs{x}, t)$. The time step between snapshots is $\Delta t_s = 0.5
\nu / u_\tau^2$, which is sufficient to resolve the characteristic
time scales in near-wall turbulence \cite{lozano2014}.

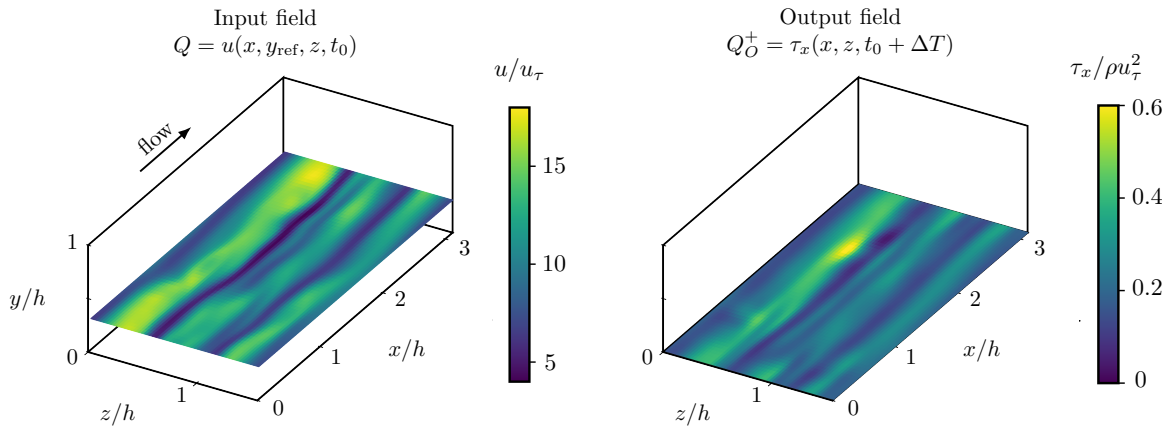
\begin{figure}
    \centering
    \hspace{-1.5cm}
    \scalebox{0.9}{
    \begin{tikzpicture}[scale=0.94, transform shape,
                        >={Latex[length=.2cm]},
                        barline/.style={thick,black!70,rounded corners=.2mm}]
      \node[inner sep=0pt, outer sep=0pt, anchor=south west] (img) at (0,0)
      {\includegraphics[width=0.4\linewidth,trim=4cm 3cm 3.5cm 3cm,clip]{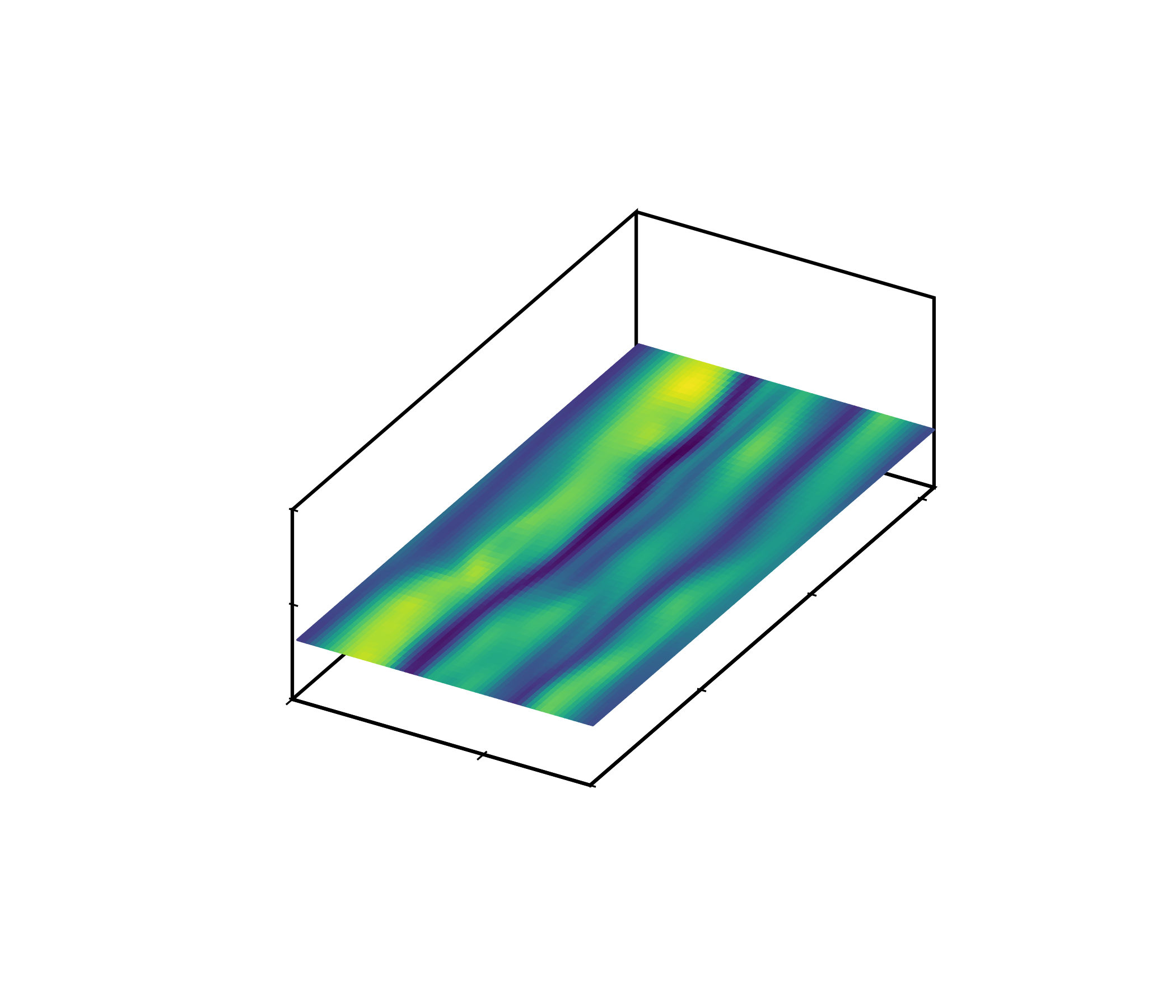}};    
      \node[above=2pt of img.north, align=center] {Input field\\[0pt] $Q = u(x,y_\text{ref},z,t_0)$};
      
      \def\xo{2}
      \def\yo{1.5}
      \def\wscale{1}
      \def\hscale{1}
    
      \def\wx{0.8}
      \def\wy{0.7}
      \def\hx{0.3}
      \def\hy{-0.1}
    
      \coordinate (P1) at (\xo,\yo);
      \coordinate (P2) at ($ (P1) + (\wscale*\wx,\wscale*\wy) $);
      \coordinate (P3) at ($ (P2) + (\hscale*\hx,\hscale*\hy) $);
      \coordinate (P4) at ($ (P1) + (\hscale*\hx,\hscale*\hy) $);

      \node at (-.7,1.8) {$y/h$};\node at (.0,0.9) {$0$};\node at (-.0,2.7) {$1$};
      \node at (.7,-0.) {$z/h$};\node at (1.7,0.3) {$1$};
      \node at (5.2,1.) {$x/h$};\node at (3.2,0.1) {$0$};\node at (4.1,0.9) {$1$};\node at (5.1,1.8) {$2$}; \node at (6.1,2.6) {$3$};

      \coordinate (Astart) at ($ (1.05,3.8) $);  
      \coordinate (Aend) at ($ (Astart) + (1*\wx,1*\wy) $);
    
      \draw[->, thick, black] (Astart) -- node[above, sloped] {flow} (Aend);
  
    \end{tikzpicture}
    \begin{minipage}{0.049\linewidth}
        \vspace{-5.4cm}
        \begin{tikzpicture}
        \node[inner sep=0pt, outer sep=0pt, anchor=south west] (img) at (0,0)
      {\includegraphics[width=\linewidth,trim=25.7cm 0.cm 1.25cm 0.cm,clip]{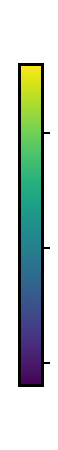}};
      \node at (0.9,4.0) {$15$};
      \node at (0.9,2.55) {$10$};
      \node at (0.85,1.1) {$5$};
      \node at (0.4,5.4) {$u/u_\tau$};
    \end{tikzpicture}
    \end{minipage}
    \hspace{1.2cm} 
    \begin{tikzpicture}[scale=0.94, transform shape]
      \node[inner sep=0pt, outer sep=0pt, anchor=south west] (img) at (0,0)
      {\includegraphics[width=0.4\linewidth,trim=4cm 3cm 3.5cm 3cm,clip]{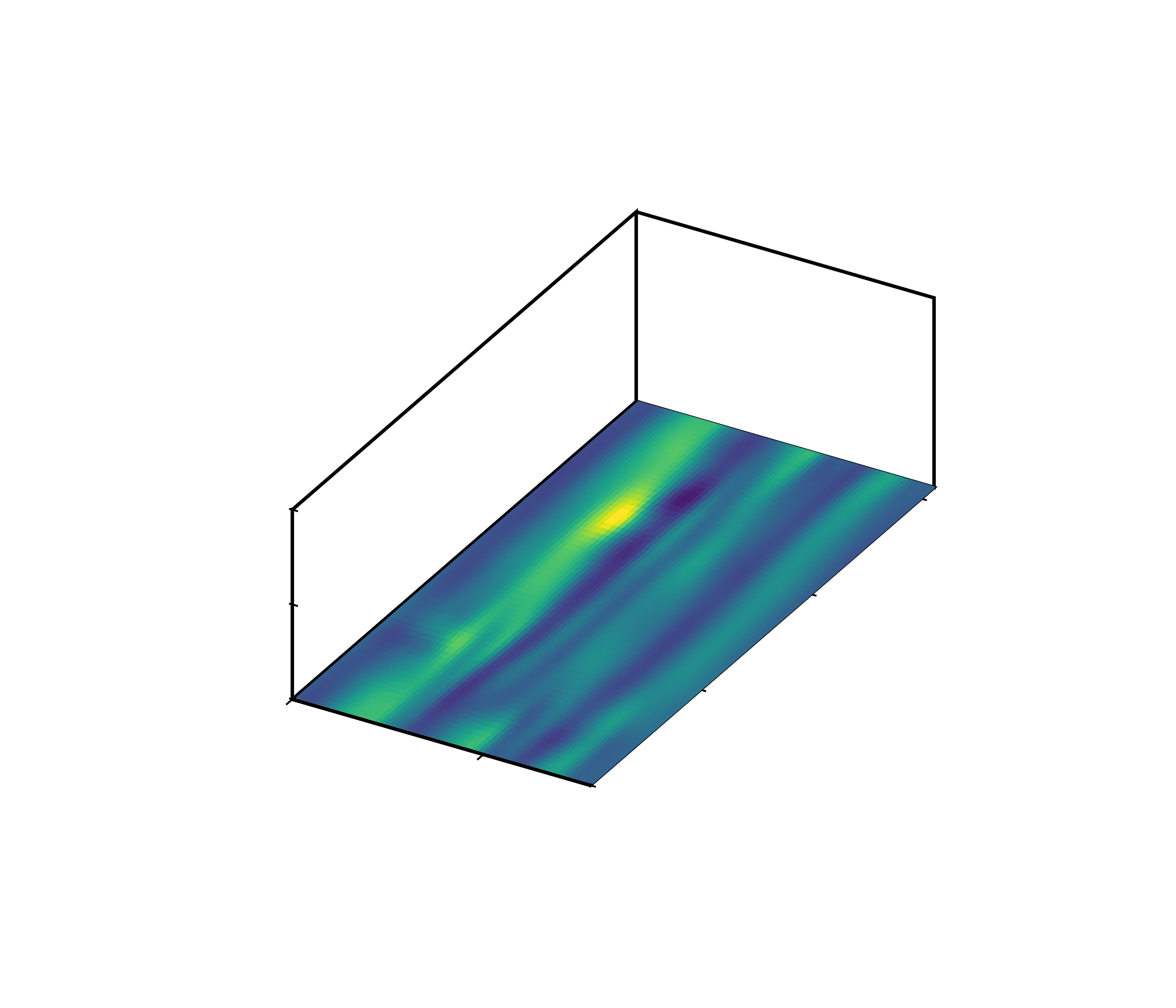}};    
      \node[above=2pt of img.north, align=center] {Output field\\[0pt]$Q_O^+ = \tau_x(x,z,t_0+\Delta T)$};
      \node at (.7,-0.) {$z/h$};\node at (1.7,0.3) {$1$}; \node at (.0,0.9) {$0$};
      \node at (5.2,1.) {$x/h$};\node at (3.2,0.1) {$0$};\node at (4.1,0.9) {$1$};\node at (5.1,1.8) {$2$}; \node at (6.1,2.6) {$3$};
    \end{tikzpicture}
    \hspace{-0.1cm}
    \begin{minipage}{0.0475\linewidth}
        \vspace{-5.4cm}
        \begin{tikzpicture}
        \node[inner sep=0pt, outer sep=0pt, anchor=south west] (img) at (0,0)
      {\includegraphics[width=\linewidth,trim=25.5cm 0.cm 1.5cm 0.cm,clip]{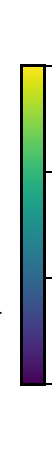}};
      \node at (1.,4.9) {$0.6$};
      \node at (1.,3.6) {$0.4$};
      \node at (1.,2.25) {$0.2$};
      \node at (0.9,0.9) {$0$};
      \node at (0.4,5.5) {$\tau_x/\rho u_\tau ^ 2$};
    \end{tikzpicture}
    \end{minipage}}
    \caption{
    Illustration of the input and output fields used for causal analysis. The left panel shows the input field $Q = u(x, y_\text{ref}, z, t_0)$, which corresponds to the streamwise velocity at a fixed wall-normal location $y_\text{ref}$ and time $t_0$. The right panel displays the output field $Q_O^+ = \tau_x(x, z, t_0 + \Delta T)$, representing the future streamwise wall-shear stress at a time lag $\Delta T$.}
    \label{fig:snap-3d}
\end{figure}

Depending on the case under consideration, the input to our analysis
consists of one or more velocity fluctuation components extracted at
selected wall-normal locations. The output is the streamwise or
spanwise wall-shear stress at a future time. For instance, for the
streamwise component:
\begin{equation}
\tau_x^+\equiv \tau_x(x, z, t + \Delta T) = \rho \nu \left. \frac{\partial u(\bs{x}, t + \Delta T)}{\partial y} \right|_{y=0},
\end{equation}
where $\rho$ is the fluid density, $u$ is the instantaneous streamwise
velocity, and $\Delta T$ is the prediction
horizon. Figure~\ref{fig:snap-3d} shows representative examples of the
input and output fields used in the analysis. The input corresponds to
a slice of the streamwise velocity fluctuations at a given wall-normal
location $y_{\rm{ref}}$ and time $t_0$, while the output is the
wall-shear stress field at the wall at time $t_0 + \Delta T$.

Our objective is to investigate the predictive capability of
forecasting models for the future wall-shear stress, while analyzing
the redundancies and synergies arising from different combinations of
wall-normal locations and velocity components. To this end, we apply
SURD to decompose the mutual information between candidate inputs and
the output into unique, redundant, and synergistic contributions. This
causal decomposition reveals combinations of input planes and
components that provide non-redundant predictive value, which guides
the optimal selection of variables in our forecasting models of the
wall-shear stress.

For comparison, we also evaluate the results obtained from SURD
against a standard space–time correlation analysis.  The correlation
between the streamwise velocity $u_i$ at a given wall-normal distance
$y_i$ and the wall-shear stress $\tau_x^+$ is defined as:
\begin{equation}
\label{eq:corr}
    C_{u_i,\tau_x^+} =
    \frac{\left|\mathbb{E}\left[(\tau_{x}^+-\mu_{\tau}) (u_i-\mu_{u})\right]\right| }{\sqrt{\mathbb{E}\left[(\tau_x^+ - \mu_{\tau})^2\right]}
     \sqrt{\mathbb{E}\left[(u_i - \mu_{u})^2\right]}},
\end{equation}
where $\mu_{\tau}=\mathbb{E}[\tau_{x}^+]$ and $\mu_{u}=\mathbb{E}[u_i]$ denote the average of $\tau_x^+$ and $u_i$, respectively, and 
$\mathbb{E}[\cdot]$ denotes the average over all spatial locations $(x,z)$ and time snapshots $t$. By construction,
the values in Eq. \ref{eq:corr} are bounded between $[0,1]$.

\subsection{Unique causality}

\begin{figure}
    \centering
    \hspace{-0.5cm}
    \scalebox{1.25}{
    \scalebox{0.7}{
    \scalebox{0.775}{
    \large
    \begin{tikzpicture}[scale=0.95, transform shape,
                        >={Latex[length=.2cm]},
                        barline/.style={thick,black!70,rounded corners=.2mm}]
      \node[inner sep=0pt, outer sep=0pt, anchor=south west] (img) at (0,0)
      {\includegraphics[width=0.25\linewidth,trim=4cm 3cm 3.5cm 3cm,clip]{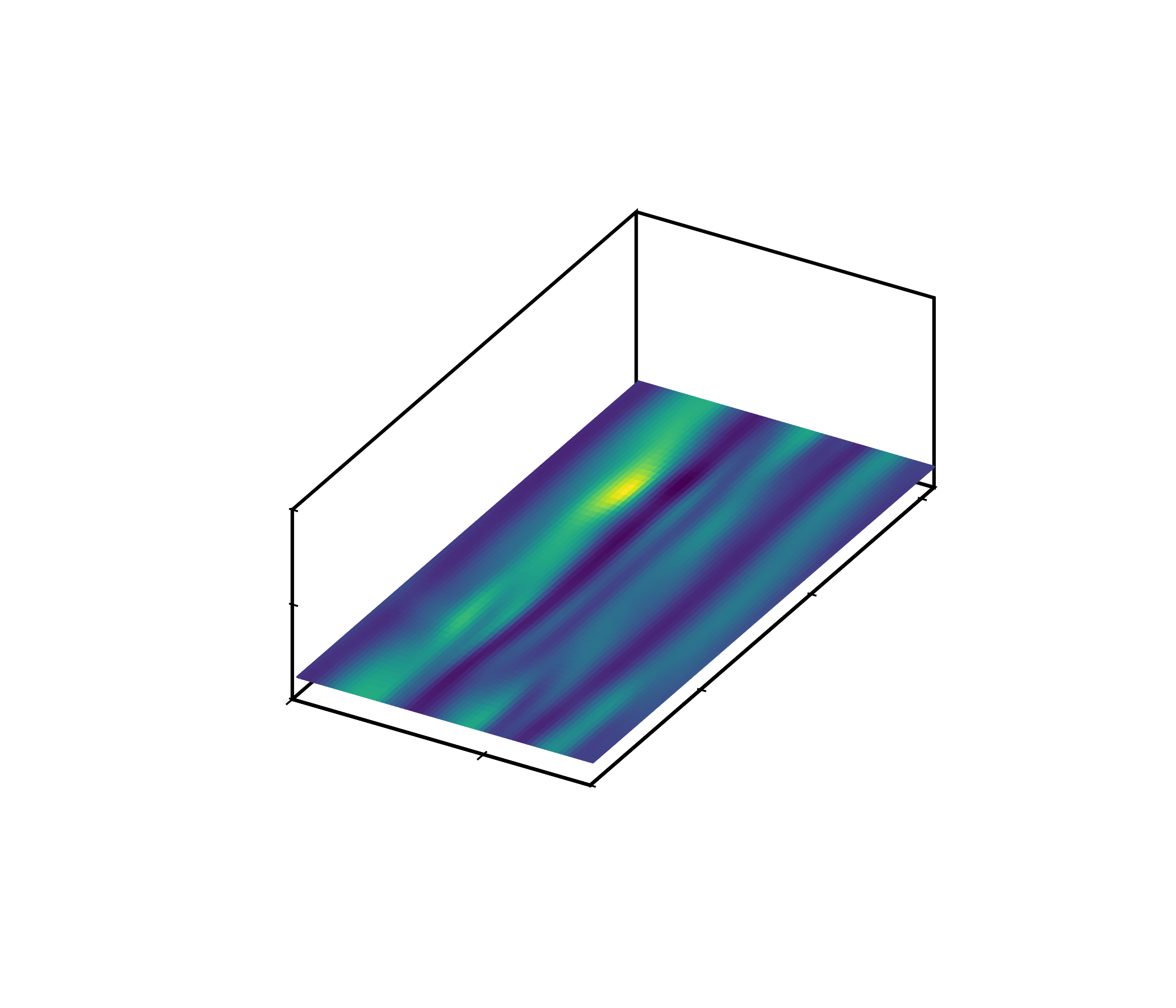}};    
      \node[above=0pt of img.north, align=center] {$u_1(\xz,t_0)$};
    
      \def\wx{0.8}
      \def\wy{0.7}
      \def\hx{0.3}
      \def\hy{-0.1}

      \node at (-.7,1.125) {$y/h$};\node at (-.05,0.5) {$0$};\node at (-.05,1.65) {$1$};
      \node at (.7,-0.2) {$z/h$};\node at (1.1,0.1) {$1$};
      \node at (2,0.0) {$0$};\node at (2.6,0.5) {$1$};\node at (3.2,1) {$2$}; \node at (3.8,1.6) {$3$};

      \coordinate (Astart) at ($ (1.05,3.8) $);  
      \coordinate (Aend) at ($ (Astart) + (1*\wx,1*\wy) $);

      \node[inner sep=0pt, outer sep=0pt, anchor=south west, rotate = 270] (img2) at (-0.45,-0.3)
      {\includegraphics[width=0.0395\linewidth,trim=25.5cm 0.cm 1.5cm 0.cm,clip]{Figures/colorbar_3d_tau.pdf}};
      \node at (-0.55,-0.65) {$u/u_\tau$};
      \node at (0.25,-1.1) {$1$};
      \node at (1.4,-1.1) {$3$};
      \node at (2.55,-1.1) {$5$};
      \node at (3.65,-1.1) {$7$};
  
    \end{tikzpicture}
    \hspace{-0.3cm}
    \begin{tikzpicture}[scale=0.95, transform shape]
      \node[inner sep=0pt, outer sep=0pt, anchor=south west] (img) at (0,0)
      {\includegraphics[width=0.25\linewidth,trim=4cm 3cm 3.5cm 3cm,clip]{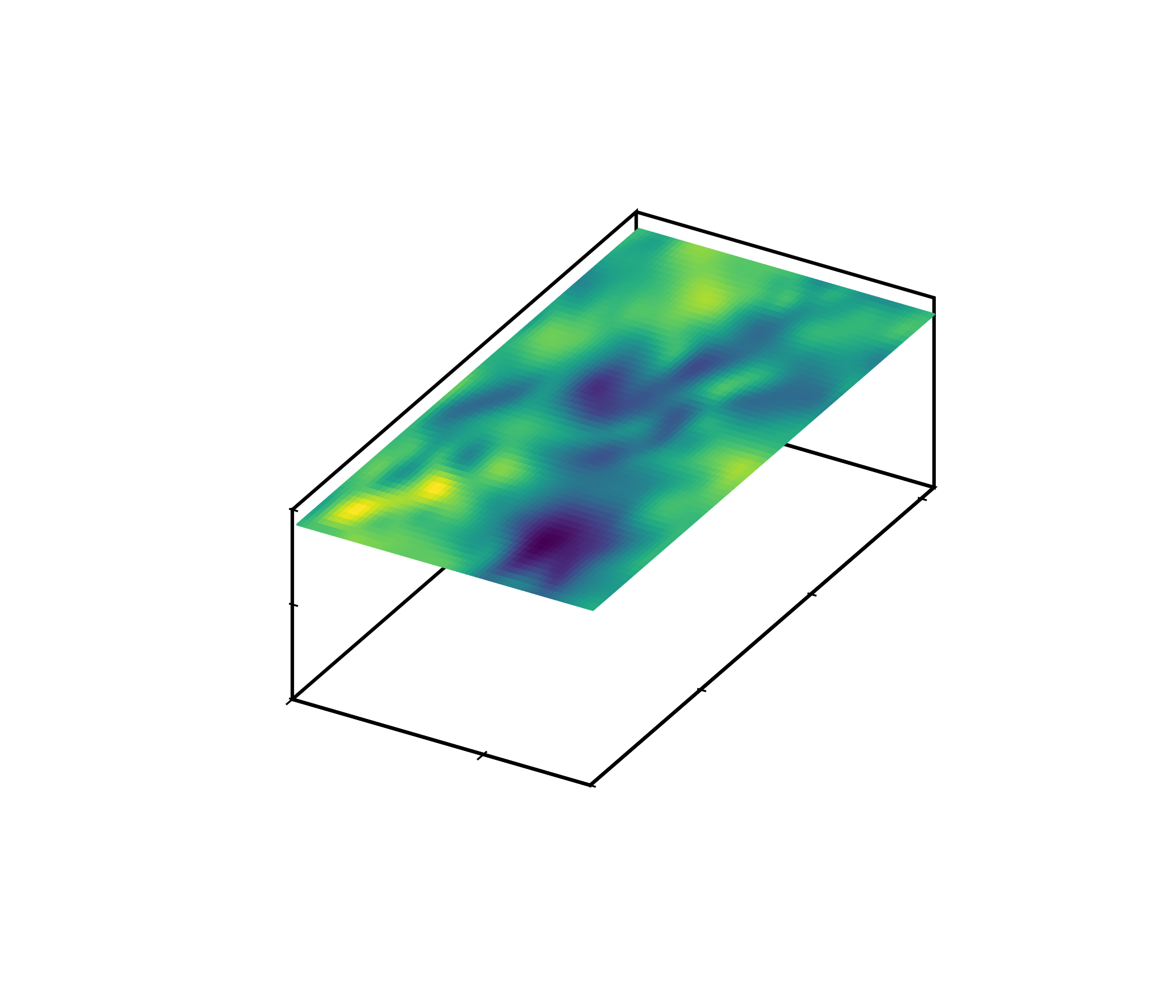}};    
      \node[above=0pt of img.north, align=center] {$u_2(\xz,t_0)$};
      \node at (.7,-0.2) {$z/h$};\node at (1.1,0.1) {$1$}; \node at (-.05,0.5) {$0$};
      \node at (3.5,0.5) {$x/h$};\node at (2,0.0) {$0$};\node at (2.6,0.5) {$1$};\node at (3.2,1) {$2$}; \node at (3.8,1.6) {$3$};

      \node[inner sep=0pt, outer sep=0pt, anchor=south west, rotate = 270] (img) at (-0.45,-0.3)
      {\includegraphics[width=0.0395\linewidth,trim=25.5cm 0.cm 1.5cm 0.cm,clip]{Figures/colorbar_3d_tau.pdf}};
      \node at (0.25,-1.1) {$17$};
      \node at (1.4,-1.1) {$18$};
      \node at (2.55,-1.1) {$19$};
      \node at (3.65,-1.1) {$20$};
    \end{tikzpicture}}
    \begin{minipage}{0.4\linewidth}
    \begin{tikzpicture}[scale=1, transform shape]
    \hspace{0.2cm}
      \node[inner sep=0pt, outer sep=0pt, anchor=south west] (img) at (0,0.8)
      {\includegraphics[width=\linewidth]{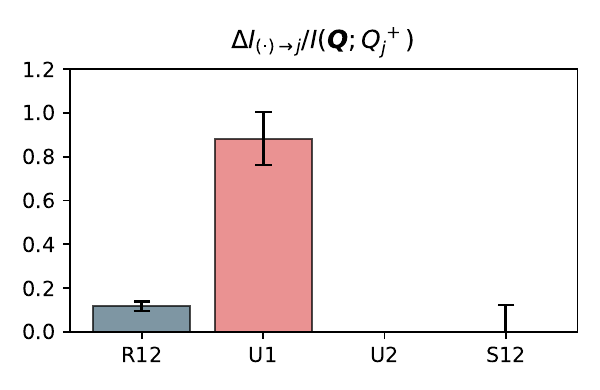}}; 
      \node[above=-17pt of img.north, align=center, fill=white] {$\quad$SURD causalities to $\tau_x^+$};
      \node[fill=white] at (0.3,1.4) {$0.0$};
      \node[fill=white] at (0.3,1.8) {$\quad$};
      \node[fill=white] at (0.3,2.3) {$0.4$};
      \node[fill=white] at (0.3,2.7) {$\quad$};
      \node[fill=white] at (0.3,3.1) {$0.8$};
      \node[fill=white] at (0.3,3.6) {$\quad$};
      \node[fill=white] at (0.3,4.0) {$1.2$};

      \node[fill=white] at (1.5,1.) {$\Delta I ^ R _ {12}$};
      \node[fill=white] at (2.7,1.) {$\Delta I ^ U _ {1}$};
      \node[fill=white] at (3.9,1.) {$\Delta I ^ U _ {2}$};
      \node[fill=white] at (5.1,1.) {$\Delta I ^ S _ {12}$};
    \end{tikzpicture}
    \vspace{3.5cm}
    \end{minipage}}
    \scalebox{0.7}{
    \begin{minipage}{0.1975\linewidth}
    \begin{tikzpicture}[scale=1, transform shape]
    \hspace{0.2cm}
       \node[inner sep=0pt, outer sep=0pt, anchor=south west] (img) at (0,0.8)
      {\includegraphics[width=\linewidth]{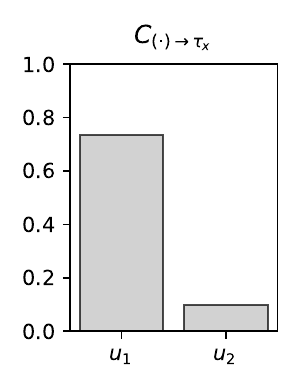}}; 
      \node[above=-17pt of img.north, align=center, fill=white] {$\quad$ $C_{(\cdot), \tau_x^+}$};
      \node[fill=white] at (0.3,1.4) {$0.0$};
      \node[fill=white] at (0.3,1.9) {$0.2$};
      \node[fill=white] at (0.3,2.425) {$0.4$};
      \node[fill=white] at (0.3,2.95) {$0.6$};
      \node[fill=white] at (0.3,3.45) {$0.8$};
      \node[fill=white] at (0.3,4.0) {$1.0$};
      \node[fill=white] at (1.3,1.) {\large$u_1$};
      \node[fill=white] at (2.3,1.) {\large$u_2$};
    \end{tikzpicture}
    \vspace{3.5cm}
    \end{minipage}
    }}
    \vspace{-3.2cm}
    \caption{ Causality between the streamwise velocity at different
      wall-normal locations and the future wall-shear stress. The left
      panels show two input fields, $u_1(\xz,t_0)$ and $u_2(\xz,t_0)$,
      corresponding to the streamwise velocity at two distinct
      wall-normal heights $y_1$ and $y_2$. These fields serve as
      inputs in the causal analysis. The colorbar is the same as in
      Figure \ref{fig:snap-3d}. The middle panel shows the resulting
      SURD causalities between these inputs and the future streamwise
      wall-shear stress $\tau_x(\xz,t_0 + \Delta T)$. The bars labeled
      $\Delta I ^ R _ {12}$, $\Delta I ^ U _ {1}$, $\Delta I ^ U _ {2}$, and $\Delta I ^ S _ {12}$ correspond to redundant, unique, and
      synergistic causal contributions from the two input layers. The
      error bars represent the variance of causalities, computed from
      100 random subsets each containing 20\% of the total data. The
      right panel shows the results of the correlation analysis using
      each input.}
    \label{fig:surd-unique}
\end{figure}

The first case analyzed consists of the analysis of the predictive
value of the streamwise velocity fluctuations $u(\xz, t)$ at two
distinct wall-normal locations for predicting the future streamwise
wall-shear stress $\tau_x(\xz, t + \Delta T)$, where $\xz = [x,z]$
denotes the spatial coordinates parallel to the wall. Specifically, we
consider two input planes: one located near the wall at $y_1^* = 5$,
within the viscous sublayer, and another located farther away, in the
center of the channel ($y_2/h=1$). Here, the superscript $(\cdot)^*$
denotes normalization in viscous units, defined as $y^* = y u_\tau /
\nu$, and should not be confused with the superscript $(\cdot)^+$,
which indicates a variable at a future time. An instantaneous
visualization of these two inputs planes is shown in Figure
\ref{fig:surd-unique}, where we refer to the streamwise velocity as as
$u_i=u(x,y_i,z,t)$.

The prediction time horizon for the future wall-shear stress is set to
$\Delta T^* = 20$, which corresponds to the moment at which it becomes
approximately independent from its own past \cite{arranz2024}. This
ensures that the predictive signal must come from external sources
rather than from the past history of the target.

Given these flow variables, we quantify the individual and joint
causal contributions from the two input planes to the future
wall-shear stress using our high-dimensional mutual information
estimation approach in combination with SURD. The results are shown in
Figure~\ref{fig:surd-unique}, where the redundant, unique, and
synergistic causal components to $\tau_x(\xz, t+\Delta T)$ are shown
in blue, red, and yellow, respectively. 

We observe that $u_1$ contain significant unique information $\Delta I
^U_{1\to \tau_x^+}$ about the output, while $u_2$ does not provide any
new information beyond what is already captured by the near-wall
input. This is evidenced by the nonzero redundant contribution $\Delta
I^R _{12\to \tau_x^+}$, the negligible unique term for the far-wall
plane $\Delta I^U _{2\to \tau_x^+}$, and the zero synergistic
contribution $\Delta I^S _{12\to \tau_x^+}$.  This implies that the
unique causal contribution from $u_1$ provides the most relevant
information for forecasting model of $\tau_x$ constructed from $u_1$
and $u_2$.

The conclusions obtained using the correlation-based approach are
similar. However, correlations do not reveal that the information in
$u_2$ about $\tau_x^+$ is redundant with that of $u_1$. Therefore,
from the perspective of feature selection, this limitation can be
misleading: correlation analysis alone might suggest that $u_2$
contributes additional information beyond $u_1$, when in fact SURD
identifies this information as redundant.
\begin{figure}[t!]
    \centering
    \begin{tikzpicture}
        \node[anchor=south west, inner sep=0] (img) at (0,0) {
            \includegraphics[width=\linewidth, trim=70pt 30pt 20pt 10pt, clip]{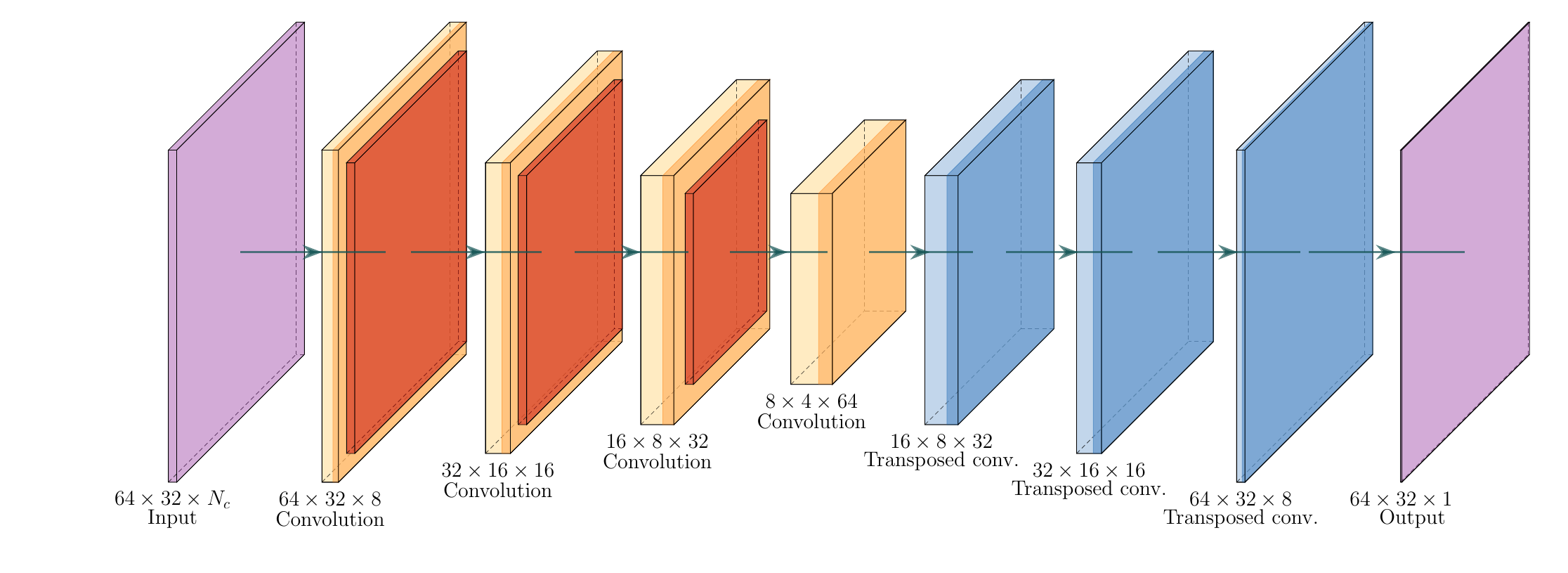}
        };
    \end{tikzpicture}
    \caption{Schematic of the architecture for CNN used for prediction
      of the wall-shear stress. The numbers below the convolution
      block denote the size of the filter and the number of channel
      applied at each layer. $N_c$ represents the number of channels
      of the input layer, which denotes the number of input variables
      $\bQ$ used for prediction of the output variable $Q_O^+$.}
    \label{fig:cnn-pred}
\end{figure}

\begin{figure}
    \centering
    \scalebox{0.95}{
    \hspace{-0.4cm}
     \begin{tikzpicture}[scale=1, transform shape,
                        >={Latex[length=.2cm]},
                        barline/.style={thick,black!70,rounded corners=.2mm}]
      \node[inner sep=0pt, outer sep=0pt, anchor=south west] (img) at (0,0)
      {\includegraphics[width=\linewidth,trim=0.5cm 0.75cm 0.5cm 2cm,clip]{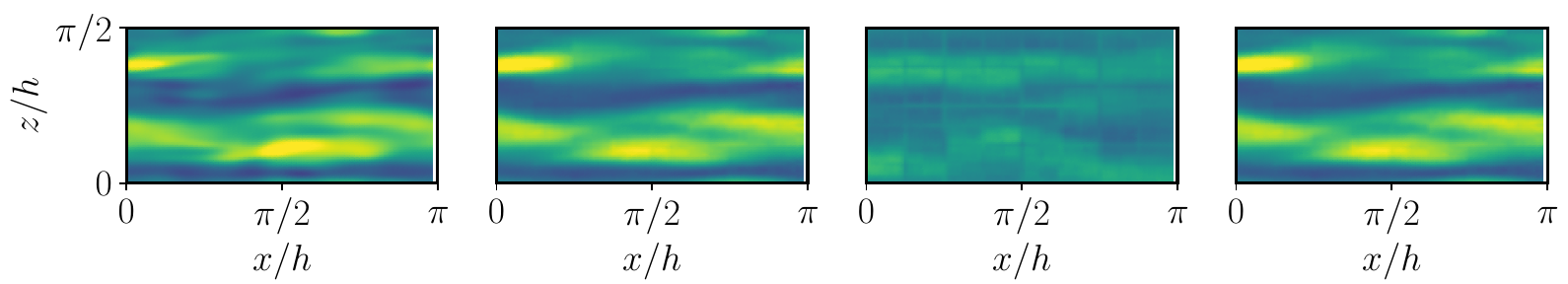}};    
      \node at (2.8,3.3) {$\tau_x^{\rm{DNS}}(\xz,t+\Delta T)$};
      \node at (6.3,3.3) {${\rm{CNN}}\left[u_1(\xz,t)\right]$};
      \node at (6.3,2.8) {${\rm{RMSE}}=0.21$};
      \node at (9.8,3.3) {${\rm{CNN}}\left[u_2(\xz,t)\right]$};
      \node at (9.8,2.8) {${\rm{RMSE}}=0.83$};
      \node at (13.3,3.3) {${\rm{CNN}}\left[u_1(\xz,t),u_2(\xz,t)\right]$};
      \node at (13.3,2.8) {${\rm{RMSE}}=0.21$};
      \node[inner sep=0pt, outer sep=0pt, anchor=south west] (img) at (15,0.175)
      {\includegraphics[width=0.025\linewidth,trim=25.5cm 0.cm 1.5cm 0.cm,clip]{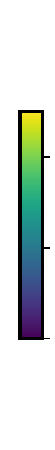}};
      \node at (15.7,2.15) {$0.4$};
      \node at (15.7,1.52) {$0.2$};
      \node at (15.57,0.95) {$0$};
      \node at (15.4,0.28) {$\tau_x/\rho u_\tau^2$};
    \end{tikzpicture}}
    \caption{ Predictions of CNN models trained with different input
      combinations of the streamwise velocity $u_i(\xz,t_0)$ to
      forecast the future wall-shear stress $\tau_x(\xz,t_0+\Delta
      T)$.  The leftmost panel shows the ground-truth target field
      obtained from DNS.  The second and third panels show CNN
      predictions using only $u_1$ and only $u_2$, with corresponding
      relative mean-squared errors (RMSE) of 0.21 and 0.83,
      respectively.  The rightmost panel shows the prediction obtained
      when both $u_1$ and $u_2$ are used as inputs, with an RMSE of
      0.21.  }
    \label{fig:pred-unique}
\end{figure}
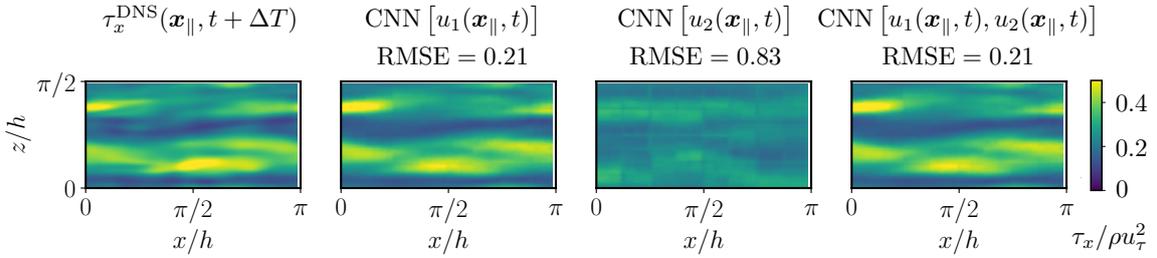

To explore the practical implications of these interactions, we train
three additional convolutional neural network (CNN) models to predict
the future wall-shear stress $\tau_x(\xz, t + \Delta T)$ using the
past streamwise velocity fluctuations as input. A schematic of the CNN
architecture used in these new predictive models is shown in
Figure~\ref{fig:cnn-pred}. The network consists of a sequence of
convolutional layers with progressively increasing channel depth and
decreasing spatial resolution, followed by an upsampling decoder that
reconstructs the output at the original resolution from different
combinations of inputs. The number of input channels $N_c$ depends on
the number of velocity planes used as input. For example, $N_c = 1$
when using only $u_1$ or $u_2$, and $N_c = 2$ when using both.

The prediction results are shown in Figure \ref{fig:pred-unique}. The
leftmost panel presents the ground-truth wall-shear stress from the
DNS data, followed by CNN predictions obtained with different
combinations of the input fields $u_1$ and $u_2$. Model performance is
quantified using the relative mean-squared error (RMSE), defined as:
\begin{equation}
    \mathrm{RMSE} = \frac{\langle\left[ \hat{\tau}_{x} - \tau_{x}^{\text{DNS}} \right]^2\rangle_{x,z,t}}{ \langle\left[\tau_{x}^{\text{DNS}}\right]^2\rangle_{x,z,t}},
\end{equation}
where $\hat{\tau}_{x}$ and $\tau_{x}^{\text{DNS}}$ represent the
predicted and reference wall-shear stress fields, respectively, and  $\langle\cdot\rangle_{x,z,t}$ denotes the average over all spatial locations $(x,z)$ and time snapshots $t$.

The remaining panels in Figure~\ref{fig:pred-unique} show the
predictions from three models: ${\rm{CNN}}\left[u_1(\xz,t)\right]$,
${\rm{CNN}}\left[u_2(\xz,t)\right]$, and ${\rm{CNN}}\left[u_1(\xz,t),
  u_2(\xz,t)\right]$, which achieve RMSE values of 0.21, 0.83, and
0.21, respectively. The best performance is obtained using only the
near-wall input $u_1$, consistent with its strong unique causal
contribution identified in the SURD analysis. In contrast, the model
based solely on the far-wall input $u_2$ exhibits a much higher error,
indicating that $u_2$ carries little predictive information about the
future wall-shear stress. Finally, simultaneously using $u_1$ and
$u_2$ yields no improvement over using $u_1$ alone, which confirms
that the information from the far-wall input is redundant with respect
to the near-wall information.

\subsection{Redundant causality}

\begin{figure}[t!]
    \centering
    \hspace{-0.5cm}
    \scalebox{1.25}{
    \scalebox{0.7}{
    \scalebox{0.775}{
    \large
    \begin{tikzpicture}[scale=0.95, transform shape,
                        >={Latex[length=.2cm]},
                        barline/.style={thick,black!70,rounded corners=.2mm}]
      \node[inner sep=0pt, outer sep=0pt, anchor=south west] (img) at (0,0)
      {\includegraphics[width=0.25\linewidth,trim=4cm 3cm 3.5cm 3cm,clip]{Figures/ch_re180_sanpshot_3d_u_y5.png}};    
      \node[above=0pt of img.north, align=center] {$u_1(\xz,t_0)$};
    
      \def\wx{0.8}
      \def\wy{0.7}
      \def\hx{0.3}
      \def\hy{-0.1}

      \node at (-.7,1.125) {$y/h$};\node at (-.05,0.5) {$0$};\node at (-.05,1.65) {$1$};
      \node at (.7,-0.2) {$z/h$};\node at (1.1,0.1) {$1$};
      \node at (2,0.0) {$0$};\node at (2.6,0.5) {$1$};\node at (3.2,1) {$2$}; \node at (3.8,1.6) {$3$};

      \coordinate (Astart) at ($ (1.05,3.8) $);  
      \coordinate (Aend) at ($ (Astart) + (1*\wx,1*\wy) $);

      \node[inner sep=0pt, outer sep=0pt, anchor=south west, rotate = 270] (img2) at (-0.45,-0.3)
      {\includegraphics[width=0.0395\linewidth,trim=25.5cm 0.cm 1.5cm 0.cm,clip]{Figures/colorbar_3d_tau.pdf}};
      \node at (-0.55,-0.65) {$u/u_\tau$};
      \node at (0.25,-1.1) {$1$};
      \node at (1.4,-1.1) {$3$};
      \node at (2.55,-1.1) {$5$};
      \node at (3.65,-1.1) {$7$};
  
    \end{tikzpicture}
    \hspace{-0.3cm}
    \begin{tikzpicture}[scale=0.95, transform shape]
      \node[inner sep=0pt, outer sep=0pt, anchor=south west] (img) at (0,0)
      {\includegraphics[width=0.25\linewidth,trim=4cm 3cm 3.5cm 3cm,clip]{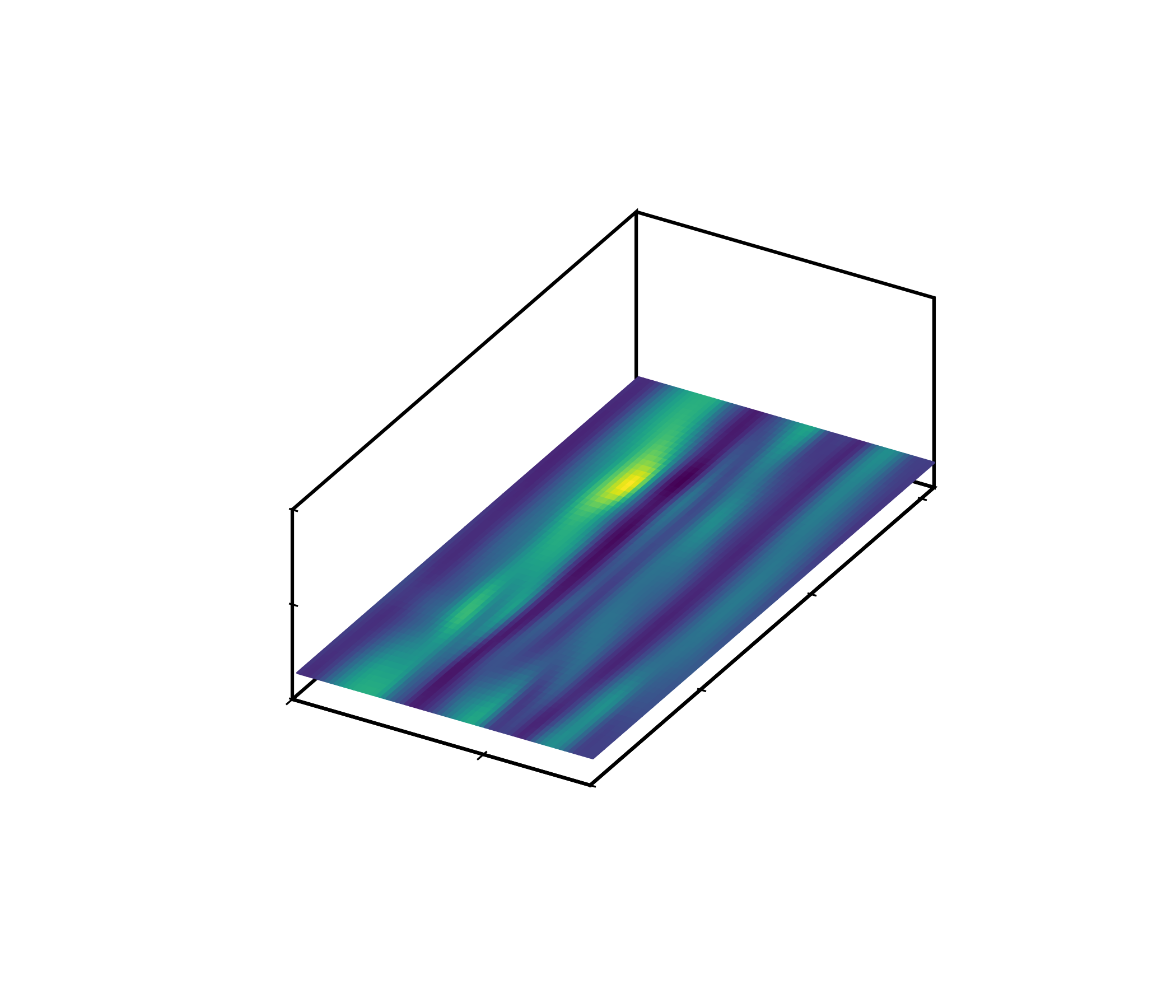}};    
      \node[above=0pt of img.north, align=center] {$u_2(\xz,t_0)$};
      \node at (.7,-0.2) {$z/h$};\node at (1.1,0.1) {$1$}; \node at (-.05,0.5) {$0$};
      \node at (3.5,0.5) {$x/h$};\node at (2,0.0) {$0$};\node at (2.6,0.5) {$1$};\node at (3.2,1) {$2$}; \node at (3.8,1.6) {$3$};

        \node[inner sep=0pt, outer sep=0pt, anchor=south west, rotate = 270] (img2) at (-0.45,-0.3)
      {\includegraphics[width=0.0395\linewidth,trim=25.5cm 0.cm 1.5cm 0.cm,clip]{Figures/colorbar_3d_tau.pdf}};
      \node at (0.25,-1.1) {$1$};
      \node at (1.4,-1.1) {$3$};
      \node at (2.55,-1.1) {$5$};
      \node at (3.65,-1.1) {$7$};
      
    \end{tikzpicture}}
    \begin{minipage}{0.4\linewidth}
    \begin{tikzpicture}[scale=1, transform shape]
    \hspace{0.2cm}
      \node[inner sep=0pt, outer sep=0pt, anchor=south west] (img) at (0,0.8)
      {\includegraphics[width=\linewidth]{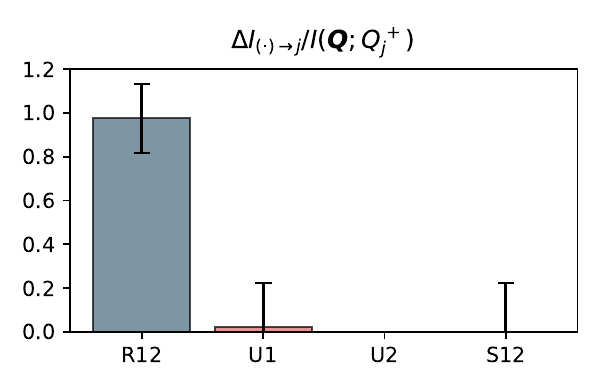}}; 
      \node[above=-17pt of img.north, align=center, fill=white] {$\quad$SURD causalities to $\tau_x^+$};
      \node[fill=white] at (0.3,1.4) {$0.0$};
      \node[fill=white] at (0.3,1.8) {$\quad$};
      \node[fill=white] at (0.3,2.3) {$0.4$};
      \node[fill=white] at (0.3,2.7) {$\quad$};
      \node[fill=white] at (0.3,3.1) {$0.8$};
      \node[fill=white] at (0.3,3.6) {$\quad$};
      \node[fill=white] at (0.3,4.0) {$1.2$};

      \node[fill=white] at (1.5,1.) {$\Delta I ^ R _ {12}$};
      \node[fill=white] at (2.7,1.) {$\Delta I ^ U _ {1}$};
      \node[fill=white] at (3.9,1.) {$\Delta I ^ U _ {2}$};
      \node[fill=white] at (5.1,1.) {$\Delta I ^ S _ {12}$};
    \end{tikzpicture}
    \vspace{3.5cm}
    \end{minipage}}
    \scalebox{0.7}{
    \begin{minipage}{0.1975\linewidth}
    \begin{tikzpicture}[scale=1, transform shape]
    \hspace{0.2cm}
       \node[inner sep=0pt, outer sep=0pt, anchor=south west] (img) at (0,0.8)
      {\includegraphics[width=\linewidth]{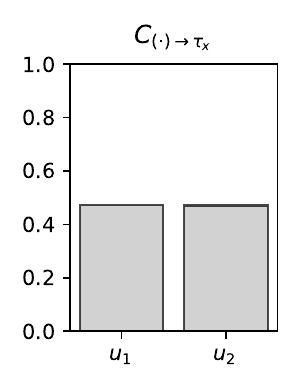}}; 
      \node[above=-17pt of img.north, align=center, fill=white] {$\quad$ $C_{(\cdot),\tau_x^+}$};
      \node[fill=white] at (0.3,1.4) {$0.0$};
      \node[fill=white] at (0.3,1.9) {$0.2$};
      \node[fill=white] at (0.3,2.425) {$0.4$};
      \node[fill=white] at (0.3,2.95) {$0.6$};
      \node[fill=white] at (0.3,3.45) {$0.8$};
      \node[fill=white] at (0.3,4.0) {$1.0$};
      \node[fill=white] at (1.3,1.) {\large$u_1$};
      \node[fill=white] at (2.3,1.) {\large$u_2$};
    \end{tikzpicture}
    \vspace{3.5cm}
    \end{minipage}
    }}
    \vspace{-3.2cm}
    \caption{ Causality between the streamwise velocity at different
      wall-normal locations and the future wall-shear stress. The left
      panels show two input fields, $u_1(\xz,t_0)$ and $u_2(\xz,t_0)$,
      corresponding to the streamwise velocity at two distinct
      wall-normal heights $y_1$ and $y_2$. These fields serve as
      inputs in the causal analysis. The colorbar is the same as in
      Figure \ref{fig:snap-3d}. The middle panel shows the resulting
      SURD causalities between these inputs and the future streamwise
      wall-shear stress $\tau_x(\xz,t_0 + \Delta T)$. The bars labeled
      $\Delta I ^ R _ {12}$, $\Delta I ^ U _ {1}$, $\Delta I ^ U _ {2}$, and $\Delta I ^ S _ {12}$ correspond to redundant, unique, and
      synergistic causal contributions from the two input layers. The
      error bars represent the variance of causalities, computed from
      100 random subsets each containing 20\% of the total data. The
      right panel shows the results of the correlation analysis using
      each input. }
    \label{fig:surd-redundant}
\end{figure}

We now consider a case where both input planes of the streamwise
velocity fluctuations are positioned close to the wall and in close
proximity to each other, at $y_1^* = 5$ and $y_2^* =
6$. Figure~\ref{fig:surd-redundant} shows an instantaneous snapshot of
the fields at these two wall-normal locations. The prediction target
remains the same as in the previous section: the future streamwise
wall-shear stress, $\tau_x(\xz, t + \Delta T)$, evaluated at a time
horizon of $\Delta T^* = 20$.

The SURD causal contributions from the two input planes are shown in
Figure~\ref{fig:surd-redundant}. Here, the dominant contribution is
the redundant term $\Delta I ^ R _ {12\to \tau_x^+}$, while the unique $\Delta I ^ U _ {1\to \tau_x^+}$, $\Delta I ^ U _ {2\to \tau_x^+}$ and synergistic
$\Delta I ^ S _ {12\to \tau_x^+}$ components remain comparatively small. This indicates that, as
expected, both planes contain mostly the same information about the
future wall-shear stress, and there is no additional value in using
them concurrently.

The correlation analysis in this case assigns nearly identical values
to $C_{u_1,\tau_x^+}$ and $C_{u_2,\tau_x^+}$.  While this outcome is
consistent with the fact that both planes carry similar information
about $\tau_x^+$, it does not indicate that this information is
redundant.  In other words, correlation analysis cannot distinguish
whether the two inputs provide overlapping content or genuinely
independent contributions.
\begin{figure}
    \centering
    \scalebox{0.95}{
    \hspace{-0.4cm}
     \begin{tikzpicture}[scale=1, transform shape,
                        >={Latex[length=.2cm]},
                        barline/.style={thick,black!70,rounded corners=.2mm}]
      \node[inner sep=0pt, outer sep=0pt, anchor=south west] (img) at (0,0)
      {\includegraphics[width=\linewidth,trim=0.5cm 0.75cm 0.5cm 2cm,clip]{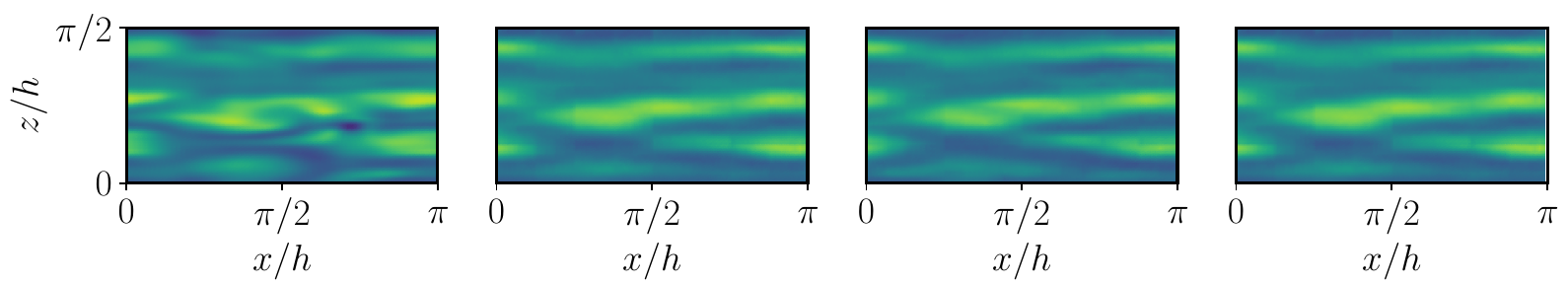}};    
      \node at (2.8,3.3) {$\tau_x^{\rm{DNS}}(\xz,t+\Delta T)$};
      \node at (6.3,3.3) {${\rm{CNN}}\left[u_1(\xz,t)\right]$};
      \node at (6.3,2.8) {${\rm{RMSE}}=0.21$};
      \node at (9.8,3.3) {${\rm{CNN}}\left[u_2(\xz,t)\right]$};
      \node at (9.8,2.8) {${\rm{RMSE}}=0.22$};
      \node at (13.3,3.3) {${\rm{CNN}}\left[u_1(\xz,t),u_2(\xz,t)\right]$};
      \node at (13.3,2.8) {${\rm{RMSE}}=0.22$};
      \node[inner sep=0pt, outer sep=0pt, anchor=south west] (img) at (15,0.175)
      {\includegraphics[width=0.025\linewidth,trim=25.5cm 0.cm 1.5cm 0.cm,clip]{Figures/colorbar_2d.pdf}};
      \node at (15.7,2.15) {$0.4$};
      \node at (15.7,1.52) {$0.2$};
      \node at (15.57,0.95) {$0$};
      \node at (15.4,0.28) {$\tau_x/\rho u_\tau^2$};
    \end{tikzpicture}}
    \caption{ Predictions of CNN models trained with different input
      combinations of the streamwise velocity $u_i(\xz,t_0)$ for
      forecasting the future wall-shear stress $\tau_x(\xz,t_0+\Delta
      T)$.  The leftmost panel shows the ground-truth target field
      from DNS, while the remaining panels show CNN predictions
      obtained using $u_1$, $u_2$, and the combined inputs $[u_1,
        u_2]$.  The RMSE for each case is indicated above the
      corresponding panel.  }
    \label{fig:pred-redundant}
\end{figure}

To illustrate how redundancy affects prediction, we train the same CNN
architectures from Figure~\ref{fig:cnn-pred} with different
combinations of inputs. The results, shown in
Figure~\ref{fig:pred-redundant}, correspond to the models
${\rm{CNN}}\left[u_1(\xz,t)\right]$,
${\rm{CNN}}\left[u_2(\xz,t)\right]$, and ${\rm{CNN}}\left[u_1(\xz,t),
  u_2(\xz,t)\right]$, which achieve RMSE values of 0.21, 0.22, and
0.22, respectively. In this case, the three models yield very similar
RMSE values, which indicates that both inputs provide essentially the
same predictive information about the output, and no benefit is gained
from combining them.

This outcome is consistent with the SURD decomposition: each input is
individually predictive of $\tau_x^+$, but their combination yields no
synergistic gain. Thus, the information carried by $u_2$ is redundant
with respect to that in $u_1$, and vice versa.

\subsection{Synergistic causality}

In the last case considered, we illustrate the synergistic predictive
value of the streamwise $u_1(\xz,t)$ and spanwise $w_1(\xz,t)$
components of the velocity at the wall-normal location $y_1^*=1$ for
predicting the future magnitude of the wall-shear stress vector,
$|\boldsymbol{\tau}|^+ = |\boldsymbol{\tau}|(\bs{x},t + \Delta T) =
\sqrt{{\tau_x^+}^2+{\tau_z^+}^2}$. The wall-normal planes are
intentionally positioned near the wall to better highlight the
synergistic interactions between the input fields.
\begin{figure}
    \centering
    \hspace{-0.65cm}
    \scalebox{1.225}{
    \scalebox{0.7}{
    \scalebox{0.775}{
    \large
    \begin{tikzpicture}[scale=0.95, transform shape,
                        >={Latex[length=.2cm]},
                        barline/.style={thick,black!70,rounded corners=.2mm}]
      \node[inner sep=0pt, outer sep=0pt, anchor=south west] (img) at (0,0)
      {\includegraphics[width=0.25\linewidth,trim=4cm 3cm 3.5cm 3cm,clip]{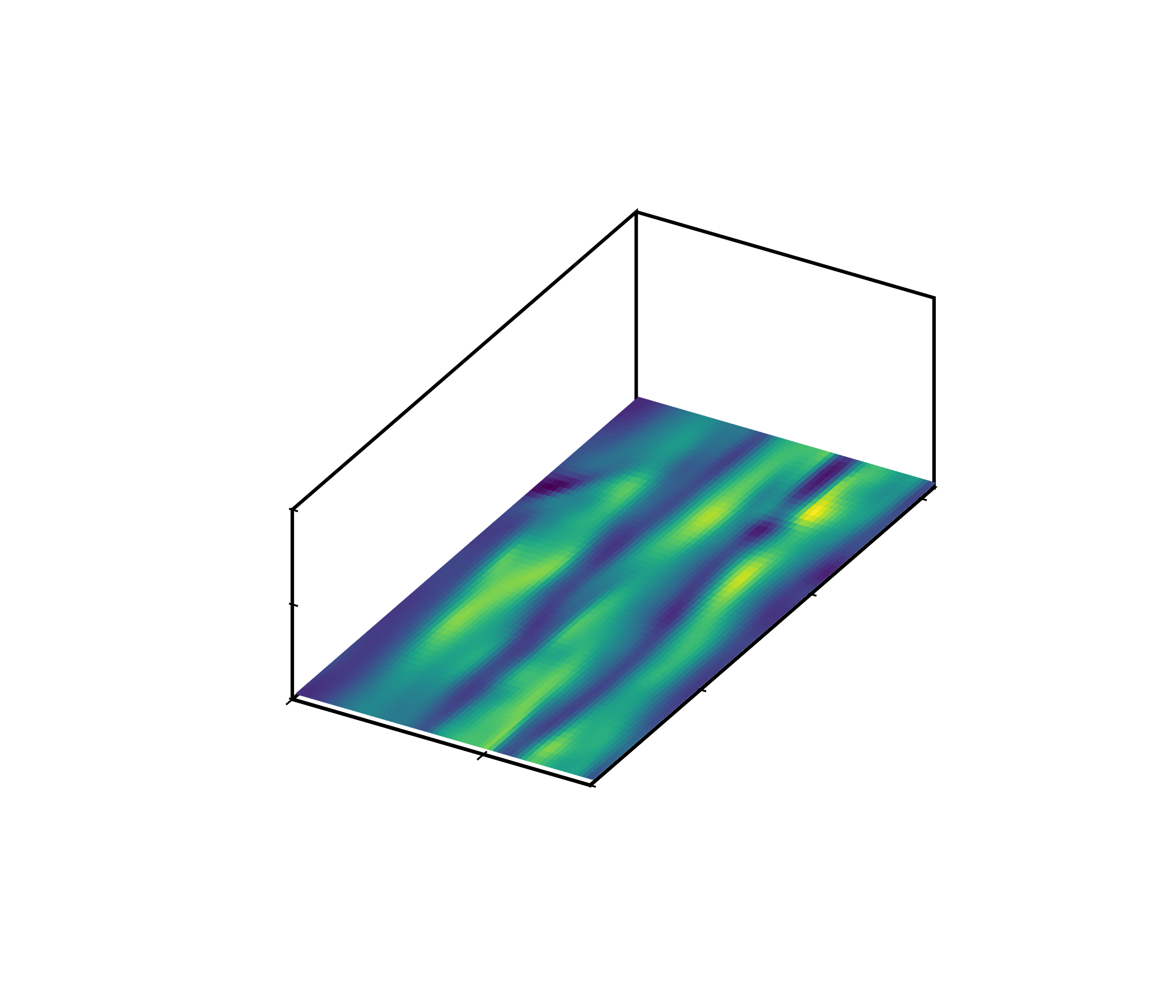}};    
      \node[above=0pt of img.north, align=center] {$u_1(\xz,t_0)$};
    
      \def\wx{0.8}
      \def\wy{0.7}
      \def\hx{0.3}
      \def\hy{-0.1}

      \node at (-.7,1.125) {$y/h$};\node at (-.05,0.5) {$0$};\node at (-.05,1.65) {$1$};
      \node at (.7,-0.2) {$z/h$};\node at (1.1,0.1) {$1$};
      \node at (2,0.0) {$0$};\node at (2.6,0.5) {$1$};\node at (3.2,1) {$2$}; \node at (3.8,1.6) {$3$};

      \coordinate (Astart) at ($ (1.05,3.8) $);  
      \coordinate (Aend) at ($ (Astart) + (1*\wx,1*\wy) $);

      \node[inner sep=0pt, outer sep=0pt, anchor=south west, rotate = 270] (img2) at (-0.45,-0.3)
      {\includegraphics[width=0.0395\linewidth,trim=25.5cm 0.cm 1.5cm 0.cm,clip]{Figures/colorbar_3d_tau.pdf}};
      \node at (-0.55,-0.65) {$u/u_\tau$};
      \node at (0.25,-1.1) {$0$};
      \node at (1.4,-1.1) {$0.2$};
      \node at (2.55,-1.1) {$0.4$};
      \node at (3.65,-1.1) {$0.6$};
  
    \end{tikzpicture}
    \hspace{-0.3cm}
    \begin{tikzpicture}[scale=0.95, transform shape]
      \node[inner sep=0pt, outer sep=0pt, anchor=south west] (img) at (0,0)
      {\includegraphics[width=0.25\linewidth,trim=4cm 3cm 3.5cm 3cm,clip]{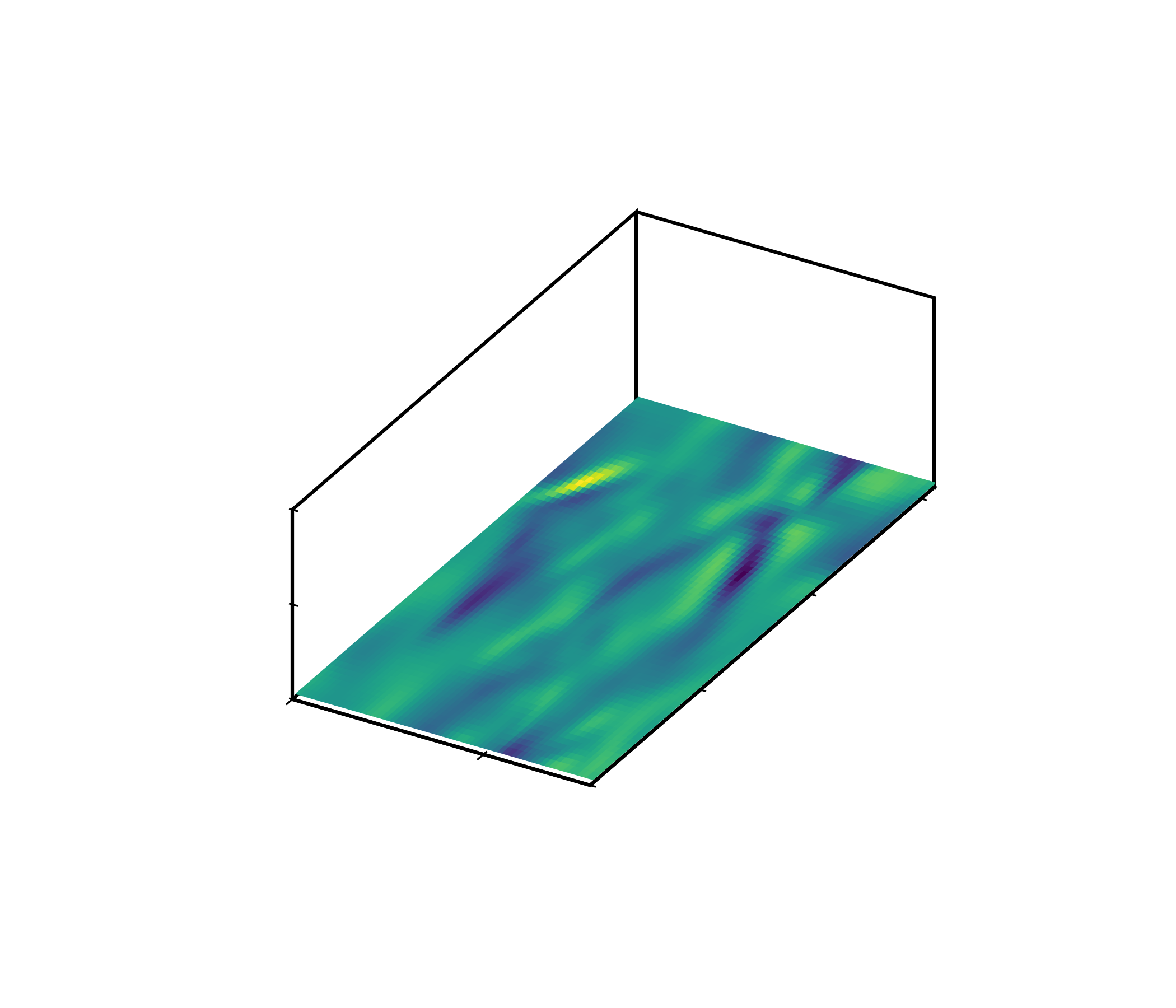}};    
      \node[above=0pt of img.north, align=center] {$w_1(\xz,t_0)$};
      \node at (.7,-0.2) {$z/h$};\node at (1.1,0.1) {$1$}; \node at (-.05,0.5) {$0$};
      \node at (3.5,0.5) {$x/h$};\node at (2,0.0) {$0$};\node at (2.6,0.5) {$1$};\node at (3.2,1) {$2$}; \node at (3.8,1.6) {$3$};
      \node[inner sep=0pt, outer sep=0pt, anchor=south west, rotate = 270] (img2) at (-0.45,-0.3)
      {\includegraphics[width=0.0395\linewidth,trim=25.5cm 0.cm 1.5cm 0.cm,clip]{Figures/colorbar_3d_tau.pdf}};
      \node at (-0.55,-0.65) {$w/u_\tau$};
      \node at (0.25,-1.1) {$0$};
      \node at (1.4,-1.1) {$0.2$};
      \node at (2.55,-1.1) {$0.4$};
      \node at (3.65,-1.1) {$0.6$};
    \end{tikzpicture}}
    \begin{minipage}{0.4\linewidth}
    \begin{tikzpicture}[scale=1, transform shape]
    \hspace{0.2cm}
      \node[inner sep=0pt, outer sep=0pt, anchor=south west] (img) at (0,0.8)
      {\includegraphics[width=\linewidth]{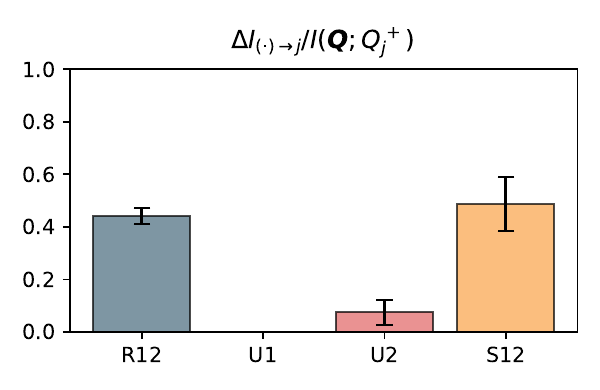}}; 
      \node[above=-17pt of img.north, align=center, fill=white] {$\quad$SURD causalities to $\vert \bs{\tau} \vert^+$};
      \node[fill=white] at (0.3,1.4) {$0.0$};
      \node[fill=white] at (0.3,1.9) {$0.2$};
      \node[fill=white] at (0.3,2.425) {$0.4$};
      \node[fill=white] at (0.3,2.95) {$0.6$};
      \node[fill=white] at (0.3,3.45) {$0.8$};
      \node[fill=white] at (0.3,4.0) {$1.0$};

      \node[fill=white] at (1.5,1.) {$\Delta I ^ R _ {12}$};
      \node[fill=white] at (2.7,1.) {$\Delta I ^ U _ {1}$};
      \node[fill=white] at (3.9,1.) {$\Delta I ^ U _ {2}$};
      \node[fill=white] at (5.1,1.) {$\Delta I ^ S _ {12}$};
    \end{tikzpicture}
    \vspace{3.5cm}
    \end{minipage}}
    \scalebox{0.7}{
    \begin{minipage}{0.1975\linewidth}
    \begin{tikzpicture}[scale=1, transform shape]
    \hspace{0.2cm}
       \node[inner sep=0pt, outer sep=0pt, anchor=south west] (img) at (0,0.8)
      {\includegraphics[width=\linewidth]{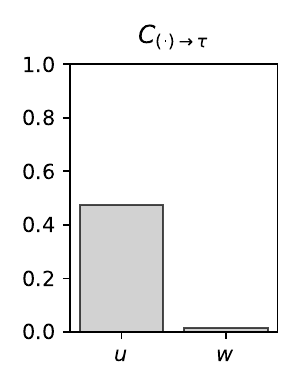}}; 
      \node[above=-17pt of img.north, align=center, fill=white] {$\quad$ $C_{(\cdot),\vert \bs{\tau} \vert^+}$};
      \node[fill=white] at (0.3,1.4) {$0.0$};
      \node[fill=white] at (0.3,1.9) {$0.2$};
      \node[fill=white] at (0.3,2.425) {$0.4$};
      \node[fill=white] at (0.3,2.95) {$0.6$};
      \node[fill=white] at (0.3,3.45) {$0.8$};
      \node[fill=white] at (0.3,4.0) {$1.0$};
      \node[fill=white] at (1.3,1.) {\large$u_1$};
      \node[fill=white] at (2.3,1.) {\large$w_1$};
    \end{tikzpicture}
    \vspace{3.5cm}
    \end{minipage}
    }}
    \vspace{-3.2cm}
    \caption{ Causality between the vector components of wall-shear
      stress and its future magnitude. The left panels show the
      streamwise and spanwise components of the velocity at the
      wall-normal height $y_1$, $u_1(\xz,t_0)$ and $w_1(\xz,t_0)$,
      respectively, used as input fields. The colorbar is the same as
      in Figure \ref{fig:snap-3d}. The middle panel displays the SURD
      causal contributions to the future magnitude of the wall-shear
      stress, $|\boldsymbol{\tau}|(\xz,t_0 + \Delta T)$. The bars labeled
      $\Delta I ^ R _ {12}$, $\Delta I ^ U _ {1}$, $\Delta I ^ U _ {2}$, and $\Delta I ^ S _ {12}$ correspond to redundant, unique, and
      synergistic causal contributions from the two input layers. The error
      bars represent the variance of the mutual information estimates,
      computed from 100 random subsets each containing 20\% of the
      total data. The right panel shows the results of the correlation
      analysis using each input.}
    \label{fig:surd-syn}
\end{figure}

The SURD causal decomposition is shown in
Figure~\ref{fig:surd-syn}. The left and center panels illustrate an
instantaneous visualization of the input fields $u_1$ and $w_1$, while
the right panel reports their causal contributions to the future
wall-shear stress magnitude $|\boldsymbol{\tau}|^+$. In this setup,
the streamwise component $u_1$ is treated as the first input and the
spanwise component $w_2$ as the second. Therefore, $\Delta I ^ U _ {1\to|\boldsymbol{\tau}|^+}$ here represents
the unique causal contribution of $u_1$ to the future of
$|\boldsymbol{\tau}|$, while $\Delta I ^ U _ {2\to|\boldsymbol{\tau}|^+}$ corresponds to that of $w_1$.

Unlike the previous cases, the dominant terms here are the synergistic
$\Delta I ^ S _ {12\to|\boldsymbol{\tau}|^+}$ and redundant $\Delta I ^ R _ {12\to|\boldsymbol{\tau}|^+}$ contributions, while the unique components
remain comparatively small. This outcome indicates that $u_1$ and
$w_1$ share some redundant information, but neither alone provides
sufficient knowledge about the future magnitude
$|\boldsymbol{\tau}|^+$. Instead, their combination yields additional
information that becomes predictive only when both are considered
together.

The correlation analysis for this configuration shows a dominant value
for $u_1$, significantly larger than that of $w_1$. While this
reflects the higher amplitude of the streamwise velocity component, it
also reveals a key limitation: correlation-based measures are strongly
influenced by the relative signal intensities rather than the true
causal contributions of the variables. As a result, one might
incorrectly infer that $u_1$ alone contains most of the predictive
information about $|\boldsymbol{\tau}|$. In contrast, the SURD
decomposition shows that both $u_1$ and $w_1$ are essential to capture
the underlying predictive structure of the future wall-shear
magnitude.

\begin{figure}
\centering
\scalebox{0.95}{
    \hspace{-0.4cm}
\begin{tikzpicture}[scale=1, transform shape,
                        >={Latex[length=.2cm]},
                        barline/.style={thick,black!70,rounded corners=.2mm}]
      \node[inner sep=0pt, outer sep=0pt, anchor=south west] (img) at (0,0)
      {\includegraphics[width=\linewidth,trim=0.5cm 0.75cm 0.5cm 2cm,clip]{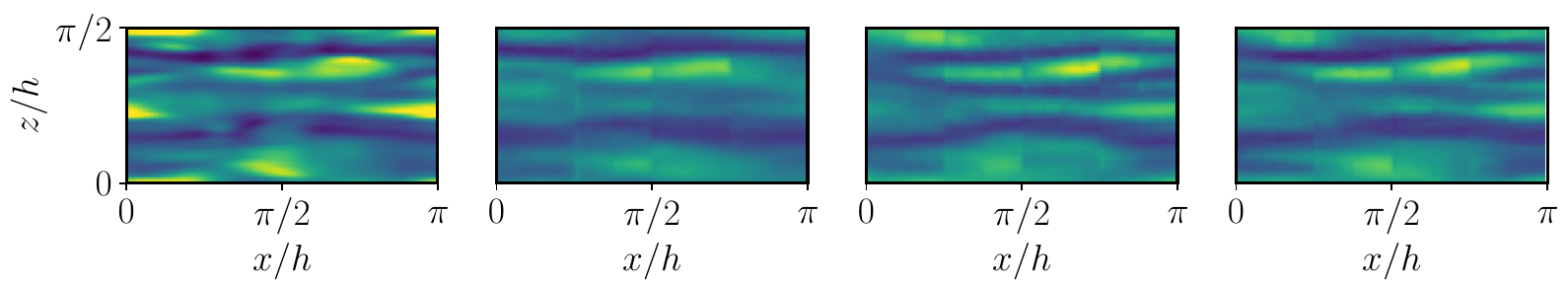}};    
      \node at (2.8,3.3) {$\vert \bs{\tau} \vert(\xz,t+\Delta T)$};
      \node at (6.3,3.3) {${\rm{CNN}}\left[u_1(\xz,t)\right]$};
      \node at (6.3,2.8) {${\rm{RMSE}}=0.32$};
      \node at (9.8,3.3) {${\rm{CNN}}\left[w_1(\xz,t)\right]$};
      \node at (9.8,2.8) {${\rm{RMSE}}=0.26$};
      \node at (13.3,3.3) {${\rm{CNN}}\left[u_1(\xz,t),w_1(\xz,t)\right]$};
      \node at (13.3,2.8) {${\rm{RMSE}}=0.19$};

      \node[inner sep=0pt, outer sep=0pt, anchor=south west] (img) at (15,0.175)
      {\includegraphics[width=0.025\linewidth,trim=25.5cm 0.cm 1.5cm 0.cm,clip]{Figures/colorbar_2d.pdf}};
      \node at (15.7,2.15) {$0.4$};
      \node at (15.7,1.52) {$0.2$};
      \node at (15.57,0.95) {$0$};
      \node at (15.4,0.28) {$\tau_x/\rho u_\tau^2$};
\end{tikzpicture}}
\caption{ Predictions of CNN models trained with different
  combinations of $u_1(\xz,t_0)$ and $w_1(\xz,t_0)$ for forecasting
  the future wall-shear stress magnitude
  $|\boldsymbol{\tau}|(\xz,t_0+\Delta T)$.  The leftmost panel shows
  the ground-truth target field from DNS, while the remaining panels
  show CNN predictions obtained using $u_1$, $w_1$, and the combined
  inputs $\left[u_1,w_1\right]$. The RMSE for each case is reported
  above the corresponding panel.  }
    \label{fig:pred-syn}
\end{figure}

To assess how this synergy affects prediction in an actual model, we
apply the same procedure as in the previous sections and train three
CNNs using different combinations of $u_1$ and $w_1$. The predictive
results are shown in Figure~\ref{fig:pred-syn}. The leftmost panel
illustrates an instantaneous visualization of the ground-truth target
from DNS, corresponding to the future wall-shear stress magnitude
$|\boldsymbol{\tau}|^+$. The subsequent panels show a visualization of
the prediction at the same time instant from the models
${\rm{CNN}}\left[u_1(\xz,t)\right]$,
${\rm{CNN}}\left[w_1(\xz,t)\right]$, and ${\rm{CNN}}\left[u_1(\xz,t),
  w_1(\xz,t)\right]$, which yield RMSE values of 0.32, 0.26, and 0.19,
respectively. These results highlight that the joint use of $u_1$ and
$w_1$ significantly improves prediction accuracy compared to either
component alone, which is consistent with the strong synergistic
contribution revealed by SURD. Thus, in this case, constructing the
most accurate predictive model of the output requires incorporating
both variables into the analysis.

\section{Discussion and conclusions}
\label{sec:discussion}

In this work, we have introduced a causality-driven approach to
analyze how synergistic, unique, and redundant interactions among
inputs constrain the fundamental limits of forecasting in chaotic
systems, independent of the specific modeling approach. This causal
characterization is achieved through the use of SURD causalities,
which enables the systematic design of minimal forecasting models that
retain only the most informative inputs while discarding those that
are irrelevant or redundant. In particular, the analysis identifies
three distinct types of contributions: inputs that offer unique
information about the output, inputs whose causal influence is
redundant with others, and inputs that contribute predictive value
only when considered jointly.

We have also shown that the combined effect of redundant, unique, and
synergistic interactions determines the minimum admissible error for
any forecasting model. This capability stems from the connection
between SURD causalities and the information-theoretic notion of
irreducible error in predictive performance. For any forecasting model
of $Q_O^+$ based on $\bQ$, the best achievable accuracy is
fundamentally constrained by the mutual information between the inputs
and the output, $I(Q_O^+; \bQ)$. The SURD decomposition exactly
recovers this quantity through its additivity property: the redundant,
unique, and synergistic components collectively sum to the total
mutual information. This information-theoretic perspective renders the
approach model-free, as the bound holds independently of the specific
algorithm or the complexity of the forecasting function class.

The results of this analysis were made possible by the use of mutual
information estimators, which allow us to approximate mutual
information in high-dimensional spaces where traditional methods are
ineffective due to the curse of dimensionality. In particular, our
approach relies on estimators based on the Donsker–Varadhan
representation, a variational method that reformulates mutual
information estimation as an optimization problem. This representation
forms the foundation of Mutual Information Neural Estimation (MINE),
which uses neural networks to learn flexible functions that
distinguish between dependent and independent variable pairs. Unlike
classical estimators that rely on discretization or density
estimation---both of which scale poorly with dimensionality---MINE
leverages the scalability of neural networks, making it well suited
for analyzing complex, high-dimensional systems such as those
encountered in turbulent flows.

The implications of this approach for designing minimal forecasting
models were demonstrated using data from a turbulent channel flow. We
first showed that isolating inputs with strong unique causal
contributions enables the construction of predictive models that
retain maximal predictive power while minimizing complexity, by
discarding variables that contribute redundant information. For
instance, when forecasting the future wall-shear stress $\tau_x(\xz, t
+ \Delta T)$, we found that the near-wall streamwise velocity field,
$u(\xz, t; y^* = 5)$, alone provides unique and sufficient predictive
information. In contrast, inputs farther from the wall (e.g., at $y/h
= 1$) offered no improvement in prediction accuracy, as their
contribution was largely redundant with that of the near-wall field.

When redundancy among input variables dominates, minimal predictive
models can be optimized by selecting a single representative variable
from the redundant set.  This interchangeability offers flexibility in
model construction, enabling variables to be chosen based on practical
factors such as ease of measurement or data availability.  In our case
study, two closely spaced near-wall fields contained duplicated
information about the future wall-shear stress, and using either field
yielded equivalent predictive performance.

In the third case analyzed, the identification of synergistic causal
contributions reveals scenarios in which no individual input variable
is sufficient on its own, but meaningful predictive information arises
from their joint interaction. In such cases, accurate forecasting
requires the inclusion of all variables participating in the
synergy. This was illustrated through the analysis of the streamwise
and spanwise velocity components, $u_1$ and $w_1$, located very close
to the wall: while neither component alone could predict the future
magnitude of the wall-shear stress, $|\boldsymbol{\tau}|$, their
combination led to a significant improvement in prediction accuracy.

\subsection{Limitations and future work}
We conclude this work by discussing some limitations of the proposed methodology.
First, the method is based on an observational definition of causality. It infers causal relationships from statistical dependencies in time-resolved data without requiring interventions. While this broadens applicability to real-world systems where interventions are infeasible or unethical, it also introduces limitations: observational causality may not coincide with interventional or counterfactual definitions of causality and can be confounded by hidden variables or latent dynamics.

Second, SURD is inherently data-intensive: accurate estimation of information-theoretic quantities in high-dimensional spaces requires large datasets, and the results can be sensitive to the choice of estimator. While the methodology itself is estimator-agnostic---valid regardless of the algorithm used---the accuracy and robustness of the results in the turbulent-flow applications may still depend on the specific mutual information estimator adopted, with potential variability across alternative estimators.

Third, SURD does not resolve the spatial or state-dependent origin of causal contributions. In the formulation used in this work, redundant, unique, and synergistic causalities are computed globally and cannot be attributed to specific flow regions or to the particular dynamical states that generate them.

Overall, we have shown that causality-driven forecasting provides an
interpretable approach for linking the underlying causal structure of
a system to its predictive performance. Future work will be devoted to
the development of methods capable of identifying the specific regions
of the flow responsible for redundant, unique, or synergistic
causalities. This follows recent works such as the Informative and
Non-informative Decomposition (IND) method proposed by
\citet{arranz2024}. Region-focused analyses of this kind would enable
a more localized understanding of causal interactions and support the
design of spatially adaptive forecasting models. In particular, when
synergy is present, it becomes especially valuable to pinpoint the
precise portions of the input fields responsible for the synergistic
effect---allowing models to retain only the informative regions while
discarding input data that do not contribute meaningfully to
prediction.

\appendix
\section{Computation of global SURD causalities}
\label{sec:app}

\begin{figure}[t!]
    \centering
    \scalebox{1}{\begin{tabular}{c}
\begin{tikzpicture}[
 			bar/.style={thick,black!70,fill=black!5,
            rounded corners=.1mm},
 			barline/.style={thick,black!70,rounded corners=.1mm},
 			myfill/.style={thick,rounded corners=.1mm},
            >={Latex[length=.2cm]},
            scale=.85
    ]
    

    \draw[thick,<->] (-.8,4.) node[anchor=north east] {$\Is$} -- (-.8,0) --+ (15.8,0);

    \def\ths{{.8,1.2,1.3,1.6,1.25,1.3,1.7 ,1.8,1.9,2.2,2., 2.4,3., 3.1,3.4}}
    \def\lbs{{4,3,1,2,34,13,24,14,23,12,234,124,134,123,1234}}

    \def\lbss{{1,2,3,4,12,13,14,23,24,34,123,124,134,234,1234}}
    \def\thss{{1.3,1.6,1.2,.8,2.2,1.3,1.8,1.9,1.7,1.25,3.1,2.4,3.1,2.,3,4}}

    \foreach \y in {0,...,14} {
        \draw[bar] (\y-.4,0) --++ (.8,0) 
        --++ (0,\thss[\y])  --++ (-.8,0) 
        -- cycle; 
        \pgfmathsetmacro{\MyPgfMathResult}{{\lbss[\y]}}
        \node[anchor=north] at (\y,0) {$\Is_{\MyPgfMathResult}$}; 
        }
    
    \draw[thin] ( 3+.5,0) --++ (0,4);
    \draw[thin] ( 9+.5,0) --++ (0,4);
    \draw[thin] (13+.5,0) --++ (0,4);


    \pgfmathsetmacro{\ypanelone}{-6};

    \draw[thick,<->] (-.8,4.+\ypanelone) node[anchor=north east] {$\Is$} -- (-.8,\ypanelone) --+ (15.8,0);

    \draw[very thin,densely dashed,black!60] (-.8,\ypanelone) --+ (15.5,0);

    \foreach \y in {0,...,14} {
        \draw[bar] (\y-.4,\ypanelone) --++ (.8,0) 
        --++ (0,\ths[\y])  --++ (-.8,0) 
        -- cycle; 
        \pgfmathsetmacro{\MyPgfMathResult}{{\lbs[\y]}}
        \node[anchor=north] at (\y,\ypanelone) {$\Is_{\MyPgfMathResult}$}; 
        }
    
    \draw[thin] ( 3+.5,\ypanelone) --++ (0,4);
    \draw[thin] ( 9+.5,\ypanelone) --++ (0,4);
    \draw[thin] (13+.5,\ypanelone) --++ (0,4);

    \draw [decorate, decoration = {brace}] ( 3.4,\ypanelone-1) --++  (-3.8,0) node [midway,anchor=north,yshift=-.5em] {$\GIs^1$};
    \draw [decorate, decoration = {brace}] ( 9.4,\ypanelone-1) --++  (-5.8,0) node [midway,anchor=north,yshift=-.5em] {$\GIs^2$};
    \draw [decorate, decoration = {brace}] (13.4,\ypanelone-1) --++  (-3.8,0) node [midway,anchor=north,yshift=-.5em] {$\GIs^3$};
    \draw [decorate, decoration = {brace}] (14.4,\ypanelone-1) --++  (-.8, 0) node [midway,anchor=north,yshift=-.5em] {$\GIs^4$};

    \pgfmathsetmacro{\ypaneltwo}{-13}

    \draw[thick,<->] (-.8,4.+\ypaneltwo) node[anchor=north east] {$\Is$} -- (-.8,\ypaneltwo) --+ (15.8,0);

    \draw[thick,<->] (-.8,4.) node[anchor=north east] {$\Is$} -- (-.8,0) --+ (15.8,0);

    \draw[barline,fill=myc1!60] (-.4,\ypaneltwo) --++ (.8,0) 
        --++ (0,\ths[0])  --++ (-.8,0) -- cycle; 
    \pgfmathsetmacro{\MyPgfMathResult}{{\lbs[0]}}
    \node[anchor=north] at (0,\ypaneltwo) {$\Is_{\MyPgfMathResult}$}; 

    \foreach \y in {1,...,14} {
        \draw[bar] (\y-.4,\ypaneltwo) --++ (.8,0) 
        --++ (0,\ths[\y])  --++ (-.8,0) 
        -- cycle; 
        \pgfmathsetmacro{\MyPgfMathResult}{{\lbs[\y]}}
        \node[anchor=north] at (\y,\ypaneltwo) {$\Is_{\MyPgfMathResult}$}; 
        }

    \foreach \y in {1,2} {
        \draw[myfill,myc1,fill=myc1!60] (\y-.4,\ths[\y-1]+\ypaneltwo) --++ (.8,0) 
        --++ (0,\ths[\y]-\ths[\y-1])  --++ (-.8,0) 
        -- cycle;
        }

    \draw[myfill,myc2,fill=myc2!60] (3-.4,\ths[2]+\ypaneltwo) --++ (.8,0) --++ 
    (0,\ths[3]-\ths[2])  --++ (-.8,0) -- cycle;

    \foreach \y in {4,5,...,9} {
        \pgfmathparse{\ths[\y] - \ths[3] > 0.0}
        \ifnum \pgfmathresult=1
            \pgfmathsetmacro{\MyPgfMathResult}{{\lbs[\y]}}
            \node[anchor=south] at (\y,\ths[\y]+\ypaneltwo) {\small$\Delta \Is^S_{\MyPgfMathResult}$};     
            \pgfmathparse{\ths[\y-1] - \ths[3] > 0.0}
            \ifnum \pgfmathresult=1
                \draw[myfill,myc3,fill=myc3!60] (\y-.4,\ths[\y-1]+\ypaneltwo) --++ (.8,0) 
                --++ (0,\ths[\y]-\ths[\y-1])  --++ (-.8,0) 
                -- cycle;
            \else
                \draw[myfill,myc3,fill=myc3!60] (\y-.4,\ths[3]+\ypaneltwo) --++ (.8,0) 
                --++ (0,\ths[\y]-\ths[3])  --++ (-.8,0) 
                -- cycle;
            \fi
        \fi
        }

    \foreach \y in {10,11,...,14} {
        \pgfmathparse{\ths[\y] - \ths[9] > 0.0}
        \ifnum \pgfmathresult=1
            \pgfmathsetmacro{\MyPgfMathResult}{{\lbs[\y]}}
            \node[anchor=south] at (\y,\ths[\y]+\ypaneltwo) {\small $\Delta \Is^S_{\MyPgfMathResult}$};     
            \pgfmathparse{\ths[\y-1] - \ths[9] > 0.0}
            \ifnum \pgfmathresult=1
                \draw[myfill,myc3,fill=myc3!60] (\y-.4,\ths[\y-1]+\ypaneltwo) --++ (.8,0) 
                --++ (0,\ths[\y]-\ths[\y-1])  --++ (-.8,0) 
                -- cycle;
            \else
                \draw[myfill,myc3,fill=myc3!60] (\y-.4,\ths[9]+\ypaneltwo) --++ (.8,0) 
                --++ (0,\ths[\y]-\ths[9])  --++ (-.8,0) 
                -- cycle;
            \fi
        \fi
        }

    \foreach \y in {0,...,14} {
        \draw[barline] (\y-.4,\ypaneltwo) --++ (.8,0) 
        --++ (0,\ths[\y])  --++ (-.8,0) 
        -- cycle; 
    }

    \node[anchor=south] at (0,\ths[0]+\ypaneltwo) {\small $\Delta \Is^R_{1234}$};
    \node[anchor=south] at (1,\ths[1]+\ypaneltwo) {\small$\Delta \Is^R_{123}$};
    \node[anchor=south] at (2,\ths[2]+\ypaneltwo) {\small$\Delta \Is^R_{12}$};
    \node[anchor=south] at (3,\ths[3]+\ypaneltwo) {\small$\Delta \Is^U_{2}$};

    \draw[densely dotted] ( 3,\ths[3]+\ypaneltwo) --++ (3, 0);
    \draw[densely dotted] ( 9,\ths[9]+\ypaneltwo) --++ (2, 0);
	
    \draw[thin] ( 3+.5,0+\ypaneltwo) --++ (0,4);
    \draw[thin] ( 9+.5,0+\ypaneltwo) --++ (0,4);
    \draw[thin] (13+.5,0+\ypaneltwo) --++ (0,4);

    \draw [decorate, decoration = {brace}] ( 3.4,\ypaneltwo-1) --++  (-3.8,0) node [midway,anchor=north,yshift=-.5em] {$\GIs^1$};
    \draw [decorate, decoration = {brace}] ( 9.4,\ypaneltwo-1) --++  (-5.8,0) node [midway,anchor=north,yshift=-.5em] {$\GIs^2$};
    \draw [decorate, decoration = {brace}] (13.4,\ypaneltwo-1) --++  (-3.8,0) node [midway,anchor=north,yshift=-.5em] {$\GIs^3$};
    \draw [decorate, decoration = {brace}] (14.4,\ypaneltwo-1) --++  (-.8, 0) node [midway,anchor=north,yshift=-.5em] {$\GIs^4$};
    




    
    \draw[very thin,white] (0,-16) --+ (1,0);

\end{tikzpicture}

\end{tabular}}
    \vspace{-0.75cm}
    \caption{Schematic of the steps involved in the calculation of causalities. The panels
  illustrate: (top) all possible mutual information values for
  a collection of four variables; (middle) tuples of mutual
  information with the components organized in ascending order; (bottom)
  the increments corresponding to redundant (blue), unique
  (red), and synergistic (yellow) causalities.}
    \label{fig:sup_method}
\end{figure}
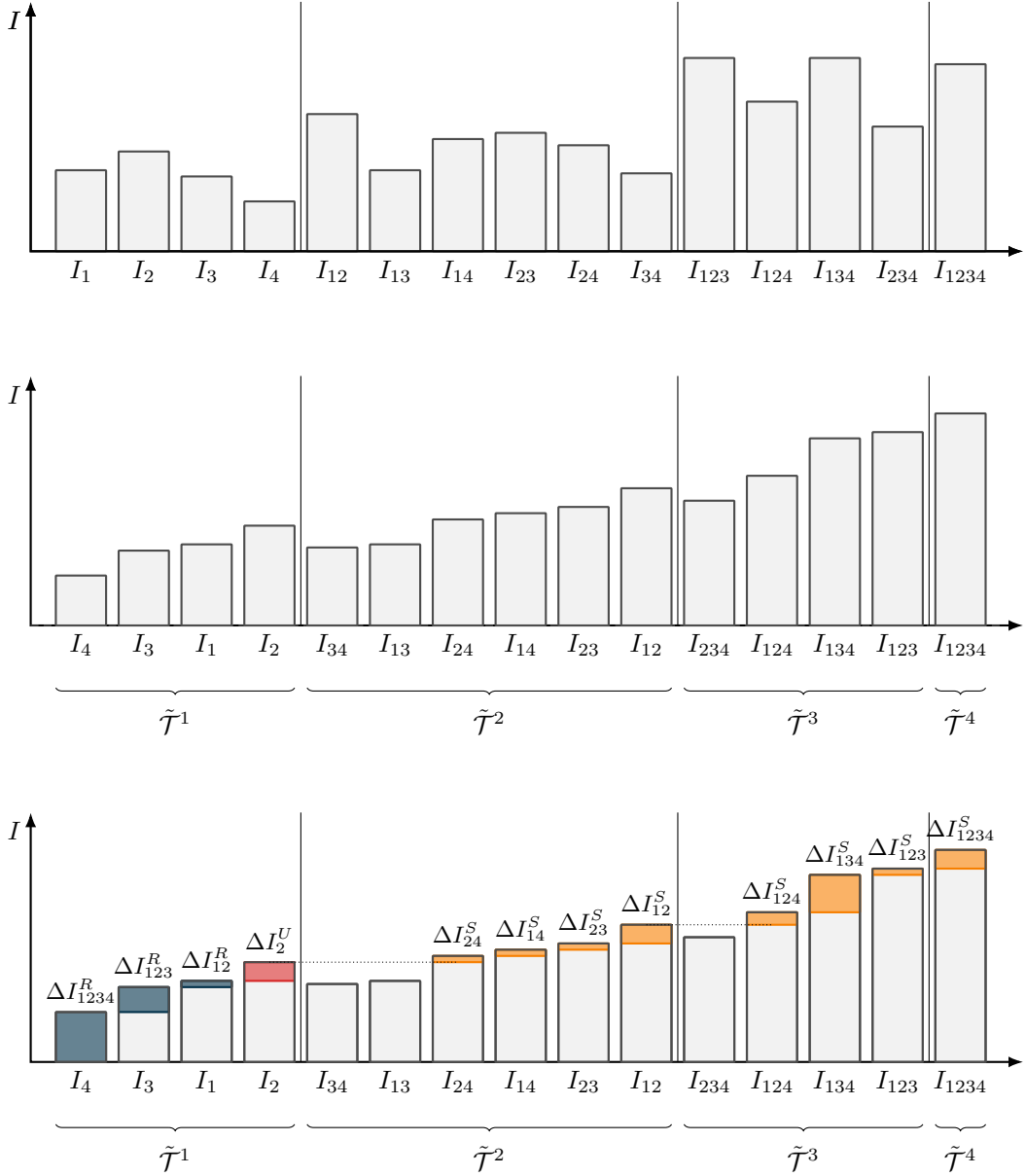

The definitions of redundant, unique, and synergistic causality adopted in this work follow the conceptual intuition outlined in \S\ref{sec:discussion}. Their computation proceeds through the following steps:
\begin{enumerate}
\item The mutual information is computed for all possible combinations
  of variables in $\bQ$ using the methodology described in
  \S\ref{sec:mine}.  This includes mutual information of order one
  ($I_1, I_2, \ldots$), order two ($\Is_{12}, \Is_{13}, \ldots$),
  order three ($\Is_{123}, \Is_{124}, \ldots$), and so on.  An
  illustrative example for a system with $N=4$ is shown in
  Figure~\ref{fig:sup_method}(a).
\item The tuples containing the mutual information of order $M$,
  denoted by $\GIs^M$, are constructed for $M=1,\ldots,N$.  The
  components of each $\GIs^M$ are organized in ascending order as
  shown in Figure~\ref{fig:sup_method}(b).
\item
  The redundant causality is the increment in information gained about
  $Q_O^+$ that is common to all the components of $\bQ_{\bj_k}$ (blue
  contributions in Figure~\ref{fig:sup_method}c):
  \begin{equation}
    \Delta \Is^R_{\bj_k} =
    \begin{cases}
      \Is_{i_k}-\Is_{i_{k-1}},& \text{for} \ \Is_{i_k},\Is_{i_{k-1}} \in \GIs^1 \ \text{and} \ k\neq n_1\\
      0,              & \text{otherwise},
    \end{cases}
  \end{equation}
  where we take $\Is_{i_0}=0$, $\bj_k = [j_{k1}, j_{k2}, \ldots]$ is
  the vector of indices satisfying $\Is_{j_{kl}} \geq \Is_{i_k}$ for
  $\Is_{j_{kl}}, \Is_{i_k} \in \GIs^1$, and $n_1$ is the number of
  elements in $\GIs^1$.
\item
  The unique causality is the increment in information gained by
  $Q_{i_k}$ about $Q_O^+$ that cannot be obtained by any other
  individual variable (red contribution in
  Figure~\ref{fig:sup_method}c):
    \begin{equation}
      \Delta \Is^U_{i_k} =
      \begin{cases}
        \Is_{i_k}-\Is_{i_{k-1}}, & \text{for}\ i_k=n_1, \ \Is_{i_{k}},\Is_{i_{k-1}} \in \GIs^1\\
        0,                 & \text{otherwise}.
      \end{cases}
    \end{equation}
    %
  \item
  The synergistic causality is the increment in information gained by
  the combined effect of all the variables in $\bQ_{\bi_k}$ that
  cannot be gained by other combination of variables $\bQ_{\bj_k}$
  (yellow contributions in Figure~\ref{fig:sup_method}c) such that
  $\Is_{\bj_k} \leq \Is_{\bi_k}$ for $\Is_{\bi_k} \in \GIs^M$ and
  $\Is_{\bj_k} \in \{\GIs^1,\ldots,\GIs^{M}\}$ with $M>1$ (dotted line
  in Figure~\ref{fig:sup_method}c):
    \begin{equation}
      \Delta \Is^S_{\bi_k} =
      \begin{cases}
        \Is_{\bi_k} - \Is_{\bi_{k-1}}, & \text{for} \ \Is_{\bi_{k-1}}\geq \max\{\GIs^{M-1}\}, \ \text{and} \ \Is_{\bi_{k}}, \Is_{\bi_{k-1}} \in \GIs^M \\
        \Is_{\bi_k} - \max\{\GIs^{M-1}\}, & \text{for} \ \Is_{\bi_{k}}>\max\{\GIs^{M-1}\}>\Is_{\bi_{k-1}}, \ \text{and} \ \Is_{\bi_{k}},  \Is_{\bi_{k-1}} \in \GIs^M\\
        0,              & \text{otherwise}.
      \end{cases}
    \end{equation}
    %
\item The redundant, unique and synergistic causalities that do not
  appear in the steps above are set to zero.
\item 
  Finally, we define the average order of causalities with respect to
  $Q_O^+$ as $N^{\alpha}_{\bi \rightarrow j}$ where $\alpha$ denotes
  R, U, or S.  The values of $N^{\alpha}_{\bi \rightarrow j}$ are used
  to plot $\Delta I^{\alpha}_{\bi \rightarrow j}$ following the order
  of appearance of $\Delta \Is^{\alpha}_{\bi \rightarrow j}$. All the
  causalities from SURD presented in this work are plotted in order
  from left to right, following $N^{\alpha}_{\bi \rightarrow j}$.
\end{enumerate}

The approach presented here differs from the original SURD formulation in that it directly uses mutual information instead of specific mutual information. The latter accounts for variations in informational contribution depending on the specific value of the output variable, $q_O^+ \in Q_O^+$. This modification enables the use of neural mutual information estimators, which efficiently approximate mutual information averaged over all states, rather than providing state-specific estimates. Nonetheless, the approach could be extended to follow the original SURD formulation by discretizing the output space and adopting a variational representation of specific mutual information---although this extension is left for future work.

\hfill

\noindent \textit{Acknowledgments} \\The authors would like to thank Gonzalo Arranz for his
contributions to this work. This work was supported by the National
Science Foundation under Grant No. 2140775 and MISTI Global Seed Funds
and UPM. Á.~M.-S. received the support of a fellowship from the "la
Caixa" Foundation (ID 100010434). The fellowship code is
LCF/BQ/EU22/11930094. The authors acknowledge the MIT SuperCloud and
Lincoln Laboratory Supercomputing Center for providing HPC resources
that have contributed to the research results reported within this
paper.\\

\noindent \textit{Disclosure statement:} \\The authors do not report potential conflicts of interest.

\printbibliography

\end{document}